\documentclass[12pt,onecolumn]{aastex63}
\usepackage{bm}
\usepackage{ulem}
\usepackage{amsmath}
\usepackage{verbatim}

\usepackage[encapsulated]{CJK}

\shortauthors{Kino et al.}
\shorttitle{ M87 jet: slowly rotating $\Omega_{F}$}

\received{April 5, 2022}
\revised{August 19, 2022}
\accepted{August 22, 2022}
\begin{document}

\title{
Implications from the velocity profile of the M87 jet:
a possibility of a slowly rotating black hole magnetosphere}

\correspondingauthor{Motoki Kino}
\email{motoki.kino@nao.ac.jp}

\author[0000-0002-2709-7338]{Motoki Kino}
\affil{Kogakuin University of Technology \& Engineering, Academic Support Center, 
 2665-1 Nakano-machi, Hachioji, Tokyo 192-0015, Japan}
\affil{National Astronomical Observatory of Japan, 2-21-1 Osawa, Mitaka, Tokyo 181-8588, Japan}

\author[0000-0001-9360-0846]{Masaaki Takahashi}
\affil{Department of Physics and Astronomy, Aichi University of Education, Kariya, Aichi 448-8542, Japan}

\author[0000-0001-8527-0496]{Tomohisa Kawashima}
\affil{Institute for Cosmic Ray Research, The University of Tokyo, 5-1-5 Kashiwanoha, Kashiwa, Chiba 277-8582, Japan}

\author[0000-0001-6558-9053]{Jongho Park}
\affil{Korea Astronomy \& Space Science Institute, Daedeokdae-ro 776, Yuseong-gu, Daejeon 34055, Republic of Korea}
\affil{
Institute of Astronomy and Astrophysics, Academia Sinica, P.O. Box 23-141, Taipei 10617, Taiwan}

\author[0000-0001-6906-772X]{Kazuhiro Hada}
\affil{Mizusawa VLBI Observatory, National Astronomical Observatory of Japan, 2-21-1 Osawa, Mitaka, Tokyo 181-8588, Japan}
\affil{Department of Astronomical Science, The Graduate University for Advanced Studies (SOKENDAI), 2-21-1 Osawa, Mitaka, Tokyo 181-8588, Japan}

\author[0000-0002-7322-6436]{Hyunwook Ro}
\affil{
Department of Astronomy, Yonsei University, Yonsei-ro 50, Seodaemun-gu, Seoul 03722, Republic of Korea}
\affil{Korea Astronomy \& Space Science Institute, Daedeokdae-ro 776, Yuseong-gu, Daejeon 34055, Republic of Korea}

\author[0000-0001-6311-4345]{Yuzhu Cui}
\affil{
Tsung-Dao Lee Institute, Shanghai Jiao Tong University, 520 Shengrong Road, Shanghai 201210, People’s Republic of China}

\begin{abstract}

Motivated by the measured velocity profile of the M87 jet using the KVN and VERA Array (KaVA) by \citet[][]{Park19} 
indicating that
the starting position of the jet acceleration is farther from the central engine of the jet than predicted in general relativistic magnetohydrodynamic simulations,
we explore how to mitigate the apparent discrepancy
between the simulations and the KaVA observation.
We use a semi-analytic jet model proposed by \citet[][]{TT03}
consistently solving the trans-magnetic field structure
but neglecting any dissipation effects.
%%%%
By comparing the jet model with the observed M87 jet velocity profile, we find that the model can
reproduce the logarithmic feature of the velocity profile, 
and can fit the observed data when choosing 
$c/(100r_{g}) \lesssim \Omega_{F} \lesssim c/(70r_{g})$
where $r_{g}$ is the gravitational radius.
%which is slower than GRMHD simulations.
%%%%%%%
While a total specific energy (${\cal E}$) of the jet 
changes the terminal bulk Lorentz factor of the jet,
a slower angular velocity of the black hole
magnetosphere (funnel region) ($\Omega_{F}$) 
makes a light-cylinder radius ($r_{\rm lc}$) larger  
and it consequently pushes out a location of a
starting point of the jet acceleration.
%%%%%%%
Using the estimated $\Omega_{F}$
we further estimate the magnetic field strength 
on the event horizon scale 
in M87 by assuming Blandford-Znajek (BZ) process is in action.
%%%
The corresponding magnetic flux threading the event horizon
of M87 is in good agreement with a magnetically arrested disc
(MAD) regime.

\end{abstract}

\section{Introduction}
\label{sec:intro}

The formation mechanism of relativistic jets  in active galactic nuclei (AGNs) remains elusive a longstanding problem in astrophysics.
Towards better understanding of this longstanding issue, 
significant forward steps have been made by recent VLBI observations of the notable nearby radio galaxy M87.
%, which motivates the present work.
The radio galaxy M87 is the one of the closest examples 
of the radio jet and 
it provides us the best opportunity to explore the jet launching mechanism at its base
\citep[e.g.,][]{Junor99, Hada11, Abramowski12, Hada14, Kim18, Walker18}.
%%%%%%
%Due to its close distance from the Earth ($D$) 
%and its large mass of the central black hole, 
%the M87 galaxy provides us the best opportunity to explore the jet launching mechanism at its base.
%%%%%%%%
Recently, the Event Horizon Telescope (EHT) has delivered the first resolved images of M87*, the supermassive black hole in the center of the M87 galaxy.
From the EHT data, the $D=16.8\pm0.8$~Mpc and
$M_{\bullet}=(6.5\pm0.7)\times 10^{9} M_{\odot}$ are derived
and the corresponding angular radius of the 
gravitational radius 
$\theta_{g}={GM_{\bullet}/Dc^{2}}=3.8\pm 0.4~{\mu}$as
where $G$ and $c$ are the gravitational constant, and
the light speed, respectively
\citep{EHT19_1,EHT19_2,EHT19_3,EHT19_4,EHT19_5,EHT19_6}.
The gravitational radius  is given by
$r_{\rm g}\equiv GM_{\bullet}/c^{2}$
and this corresponds to the relation of
$1~{\rm mas}=263~r_{g}$.

Based on mm/sub-mm VLBI observations,
the jet base of M87 has been indicated to be magnetic-energy dominated based on the energetics at the optically thick region against synchrotron self-absorption (SSA) process \citep[][]{Kino14,Kino15}.
Now
it is widely considered that magnetic field plays an important role in the formation of the relativistic jet
\citep[e.g.,][]{blandford19}.
%%%%%%%%
At the footpoint region of the jet,
a scenario in which the jet formation 
is caused by 
the extraction of the black hole's rotational energy
via the large scale magnetic field that 
penetrates the black hole event horizon, 
has been proposed by \citet[]{BZ77} (hereafter BZ77), 
which is so-called BZ process.
Although this BZ process looks promising so far, 
there are still many details of the physical process that 
are not yet understood.
Given this background, we will
discuss a jet model driven by a large-scale magnetic field
in this work.

Toward a better understanding of the jet formation,
the $z$-dependence of the jet velocity ($u_{p}$) where the subscript $p$ denotes the poloidal velocity is one of the fundamental quantities to 
be explored.
The profile of $u_{p}$ in M87 has been intensively investigated via VLBI monitoring for years
\citep{Kovalev07, Asada14, Mertens16, Hada16, Hada17, Walker18}.
%%%
Recently, further comprehensive dedicated observation of  densely-sampled monitoring 
at 22 and 43~GHz in 2016 using the KVN and VERA (KaVA) array, 
(as a sub-array of the East Asian VLBI Network
\citep[][and references therein]{Wajima16, Asada17, An18})
was performed as one of the large programs of KaVA array\citep[][]{Niinuma14, Kino15PKAS} and
it particularly clarified the velocity field on
0.3-10~mas scale by \cite{Park19}.
%%%
Interestingly, \cite{Park19} pointed out that 
the measured velocity profile is not described as a single streamline but rather 
explained by multiple ones, and 
the location where the jet starts acceleration is farther from the central engine than expected in GRMHD simulations.
The existence of such a discrepancy has been also pointed out in recent literature 
\citep{Nakamura18, Chatterjee19}.

The goal of this work is to find a possible 
solution to mitigate the apparent discrepancy between the theoretical model of magnetically accelerated jet and the observed velocity field profile in the M87 jet by
\citet{Park19}.
%%%%%
It is important to note
a possibility that blob motions observed by VLBI 
may be caused by  apparent changes of 
dissipative non-steady pattern structures such as shocks, turbulence and local instabilities \citep[e.g.,][]{Cohen14, Mertens16}
that can be different from the fluid velocity itself.
The purpose of this paper is not to deny this possibility.
%%%%%%%%
However, it is difficult to ascribe all the moving blob motions 
to dissipative pattern structures, which do not reflect actual fluid motions  
since the one-side feature  ubiquitously seen in radio jets in AGNs
is essentially explained by
the Doppler boosting effect due to actual fluid motions.
%%%
Therefore, while recognizing the possibility that some of the observed velocities could be partially mixed with pattern velocities, 
we will investigate the nature of the stationary jet in this paper.

In \S~3, we briefly overview the model proposed by
 \citet[][]{TT03} (hereafter TT03).
In \S~4, 
we show basic properties of $u_{p}$ based on TT03 model. 
In \S~5, we apply TT03 model to the M87 jet and 
we constrain on  ${\cal E}$ and $\Omega_{F}$ of M87 by
comparing the model predicted $u_{p}$ to the VLBI measured $u_{p}$.
In \S~6, we make comparisons between our result and previous works, and then discuss implications of our result.
In \S~7, we summarize the present work.
%%%
In this work, we use the natural unit ($c=1$, $G=1$), otherwise stated.

\section{Overall setting}

\begin{figure} 
\includegraphics
[width=18cm]
{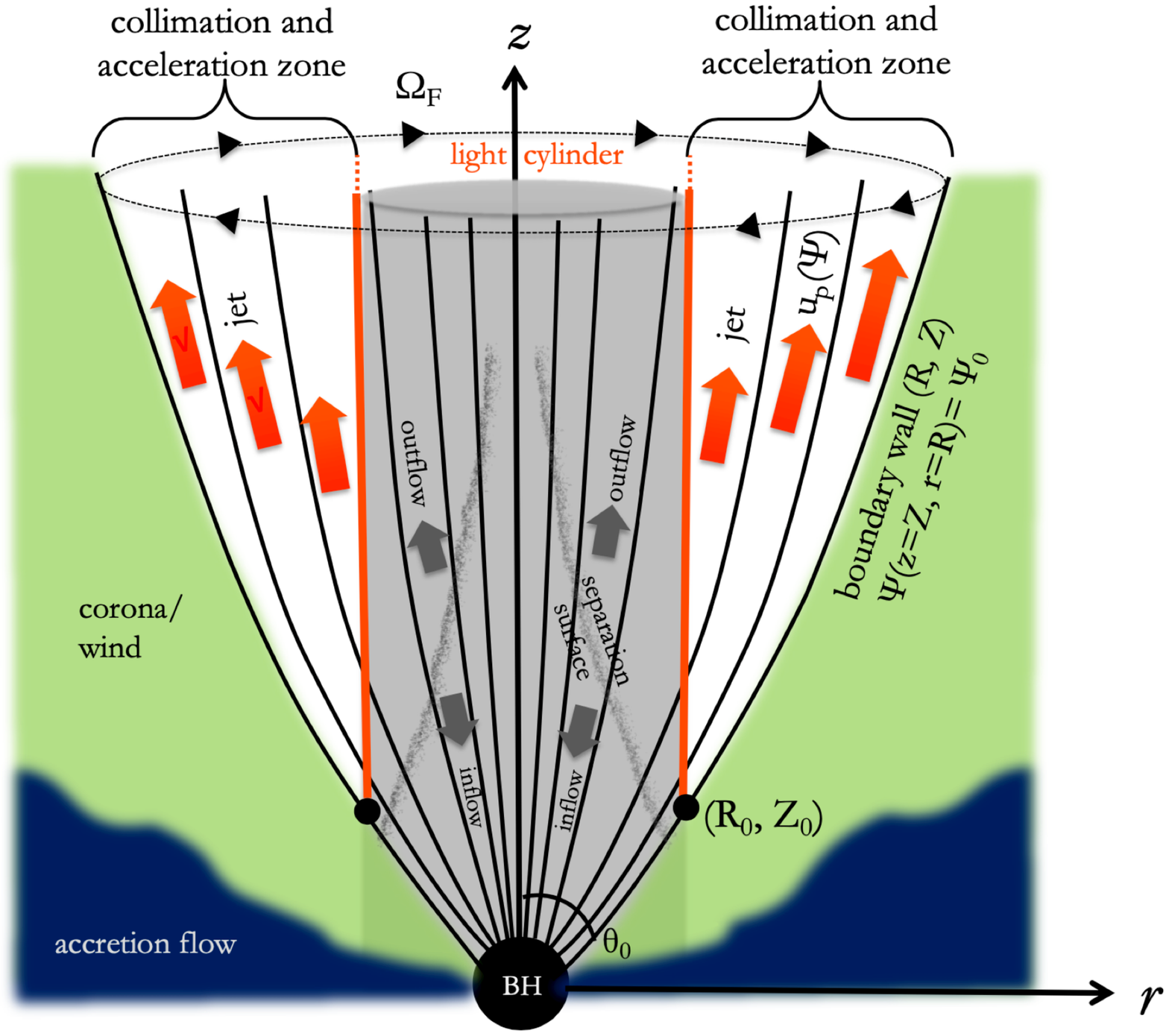}
\caption
{
%%%%
%** (technical) why is this fig updated properly??? **
%%%%
A schematic illustration of Poynting flux dominated
jet confined by the outer boundary wall made of corona/wind region.
%%%
The jet is accelerated outside the light cylinder and 
that is denoted as acceleration region (collimation and acceleration zone).
%%%
In the highly magnetized funnel region, a stagnation 
surface is generated that separates the inflow and outflow regions.
TT03 appropriately describes the poloidal velocity field of the jet flow 
in the outer region of the light cylinder (equivalent to the 
outer light surface) denoted as $u_{p}(\Psi)$. 
The angular velocity of the magnetic field lines is denoted as $\Omega_{F}$.
TT03 model does not describe the interior of the light cylinder.
We use cylindrical coordinates. 
The position coordinates of the outer boundary wall
is denoted by the capital letters ($R$, $Z$).
The value of the $z$ coordinate of the light surface is different for each magnetic field line. The position where each magnetic field line intersects the light surface is the position of the light cylinder for each magnetic field line, thus the more distant the light surface for the inner magnetic field line, the farther away it is.
We denote the $z$ coordinate where the boundary wall
(equivalent to the outermost magnetic field line) 
intersects the light surface as $(R_{0}, Z_{0})$. 
}
\label{fig:overall}
\end{figure}

Before going into a detailed description of the model, 
it would be useful to describe the overall setting 
and the core motivation of the present work.
%%%
Figure~\ref{fig:overall} summarizes
the overall picture of the situation considered here.
In this work, we utilize the MHD model proposed by 
TT03 in the framework of special relativity (i.e., SRMHD). 
It shows a schematic illustration of Poynting flux dominated
jet confined by the outer boundary wall made of corona/wind region.
%%%
In the black hole magnetosphere,
due to the balance between the gravitational force of the 
black hole and magneto-centrifugal force, a stagnation 
(also known as separation) surface is generated that separates the inflow and outflow regions
\citep[e.g.,][]{Takahashi90,McKinney06,Pu15,Pu20}.
\footnote{
TT03 model, however, does not 
include gravity. Therefore, it does not determine
the location of the separation surface.}
%%%%
For clarity, in Figure~\ref{fig:overall}, 
we show the region where we will apply TT03 model.
%%%
Comparing semi-analytical approaches and GRMHD simulation approaches, 
it is known that the semi-analytical approaches have the following advantages.
The large spatial extent of the acceleration region
has posed a challenge for such calculations by GRMHD simulations
and they tend to be eventually
limited by computational costs and numerical dissipation 
\citep[e.g.,][for details]{McKinney06, Komissarov07}, 
while semi-analytic approaches are free from these concerns.
When discussing properties of axisymmetric and steady MHD flows in general, 
the magnetic field geometry should be consistent with Grad-Shafranov (GS) equation, and the flow should be trans-fast-magnetosonic.
However, it is technically difficult to obtain a solution satisfying both of these conditions 
\citep[e.g.,][for review]{Beskin10}.
TT03 model is the only semi-analytic solution to satisfy both of these conditions.
Since $u_{p}$ profile is sensitive to the magnetic field geometry, 
we use TT03 model in this work.

TT03 model is prescribed by two model parameters, 
i.e., total energy (${\cal E}$) and the angular velocity 
 of magnetic field lines $\Omega_{F}$ for a given streamline of the flow.
Therefore, our main goal in this work can be rephrased as 
constraining ${\cal E}$ and $\Omega_{F}$ by 
matching $u_{p}$ profiles by KaVA observation and TT03 model.
%%%
It would be worth stressing in advance
that the physical quantity $\Omega_{F}$ is one of 
the most important quantities in BZ process.
%%%
The BZ process is a magnetic extraction of the spin energy of a Kerr black hole
within the force-free limit and it is thought to be a plausible production 
mechanism for the relativistic jets in AGNs.
BZ77 showed that a frame-dragging effect of the central Kerr black hole
can induce an outward flux of electromagnetic energy 
along magnetic field lines threading the event horizon,
at the expense of the black hole’s rotational energy and its expected power
($L_{BZ}$) is given by
\begin{eqnarray}
L_{BZ}\propto \Omega_{F}(\Omega_{H}- \Omega_{F}) B_{H}^{2},
\end{eqnarray}
where $B_{H}$ is 
the magnetic field strength threading the event horizon.
GRMHD simulations of jet productions indicate that powerful jets
can be produced by BZ process, when
an angular velocity of the central Kerr BH ($\Omega_{H}$) is not 
too small 
\citep[e.g.,][and references therein]{Zamaninasab14}
and higher spin of the black holes for powerful outflows 
is also in good agreement 
with the indications from the observational data
\citep[e.g.,][]{Sikora07}.
One of the questions to be addressed in this paper will be 
whether the M87 jet meets the condition of the activation of 
the BZ process, i.e., $0<\Omega_{F}<\Omega_{H}$ or not.
If the condition seems to hold in M87, 
then we will estimate $B_{H}$ 
by using the estimated  $\Omega_{F}$.

\section{Model}

We briefly overview the work of TT03.
Hereafter, the cylindrical coordinate $(t, r, \phi, z)$ is used and 
the corresponding line element is given by
$ds^{2}= c^{2} dt^{2} - dr^{2}-dz^{2}-r^{2}d\phi^{2}$.

\subsection{Basic Assumptions}

The basic assumptions in TT03 are as follows.

\begin{itemize}

\item 
A cold (zero pressure), 
steady ($\partial/\partial t =0$), axisymmetric ($\partial/\partial \phi =0$)
 special relativistic MHD jet flow is assumed.

\item

Effects of general relativity (GR) are not included 
and a Minkowski space-time is assumed in this work.
The assumption is well justified on the spatial scale  dealt with in the present work.
A formulation including GR effects 
(but assuming the magnetic field geometry)
is presented in 
\citet{Takahashi08, Pu20, Huang20}.

\item 
Any dissipation and energy loss processes are not
included in TT03 model.
Dissipation effects caused by
various instabilities are generally considered to 
become more pronounced as the jet moves downstream
\citep[e.g.,][and references therein]{Chatterjee19}.

\end{itemize}

With these assumptions, the flow is characterized by five physical quantities, i.e., 
the poloidal and toroidal velocity ($u_{p}$ and $u_{\phi}$)
and the poloidal- and toroidal magnetic field ($B_{p}$ and $B_{\phi}$), 
and the plasma mass density $\rho$ 
Poloidal magnetic field and 4-velocity of the fluid are 
given by $B_{p}^{2} = B_{z}B^{z}+B_{r}B^{r}$ and 
$u_{p}^{2} = u_{z}u^{z}+u_{r}u^{r}$, respectively.
With the Lorentz factor of the poloidal velocity 
$\Gamma=\sqrt{1+u_{p}^{2}}$,  $u_{p}=\Gamma v_{p}$ hold 
where $v_{p}$ is the 3-velocity of the poloidal velocity
\citep[][]{Tomimatsu94}.

\subsection{Field aligned conserved quantities}

Here we  introduce the well known magnetic-field aligned 
conserved quantities
(i.e.,   
${\cal E}$, 
${\cal L}$, 
$\Omega_{F}$, and 
$\eta$, see later),
which facilitates understanding flow dynamics.
%%%%%%%%%%%%%%%%%%%%%%
The $B_{p}$ and $u_{p}$ will be determined by GS equation
and  the  relativistic Bernoulli  equation together with appropriate conditions 
(such as boundary condition, trans-magnetosonic condition).
The remaining three quantities will be given by conservation laws and boundary conditions
at the plasma source.
The toroidal components $u_{\phi}$ and $B_{\phi}$ are obtained by
the $B_{p}$ and $u_{p}$ and the conserved quantities.

Using the vector potential of the magnetic field ($\vec{A}$),
the magnetic field is given by $\vec{B}=\nabla \times \vec{A}$.
A stationary and axisymmetric ideal MHD flow provides the existence of 
a magnetic flux (stream) function ($\Psi (r,z)$).
The toroidal component of $\vec{A}$
plays a role in the magnetic flux (stream) function and it is written as
\begin{eqnarray}
\Psi (r,z)= A_{\phi}
\end{eqnarray}
\citep[e.g.,][]{BZ77}. 
The magnetic fields are structured along the surface of $\Psi (r,z)=$constant.
The poloidal magnetic field is given by
 $\vec{B}_{p}=\frac{\nabla \Psi(r,z) \times \vec{e}_{\phi}}{2\pi r }$
where $\vec{e}_{\phi}$ is the $\phi$-component unit vector,
which can be written as
$B^{r}= - \frac{1}{r} \frac{\partial \Psi(r,z)}{\partial z}$, and 
$B^{z}=\frac{1}{r} \frac{\partial [\Psi(r,z)]}{\partial r}$.

For a stationary, axisymmetric MHD flow,
there are four conserved quantities along a constant $\Psi$ surface,
which are
$\eta(\Psi) $,
${\cal E}(\Psi)$, 
 ${\cal L}(\Psi)$, and
$\Omega_{F}(\Psi)$, are 
the particle flux,
the total specific energy 
the total specific angular-momentum, and
the angular velocity of a magnetic field line, respectively
\citep[][and references therein]{Camenzind86}.
%%%
%Below, we omit the mark of the argument $\Psi$ 
%in physical quantities for simplicity.
%%%
The total specific energy and angular-momentum can be
decomposed into two terms as follows.
\begin{eqnarray}
{\cal E}&=&{\cal E}_{\rm EM}+{\cal E}_{\rm MA},  \quad
 {\cal E}_{\rm EM}=-\frac{r \Omega_{F} B_{\phi}}{4\pi \eta}, \quad
 {\cal E}_{\rm MA}= \Gamma     ,
\end{eqnarray}
\begin{eqnarray}
{\cal L}={\cal L}_{\rm EM}+{\cal L}_{\rm MA}, \quad
 {\cal L}_{\rm EM}=-\frac{r B_{\phi}}{4\pi \eta}  , \quad
 {\cal L}_{\rm MA}= \Gamma r^{2}\Omega    ,
\end{eqnarray}
where $\Omega$ is the angular velocity of the plasma in the jet
and the relation $\eta= \rho u_{p}/B_{p}$ holds.
%%%
The $\sigma$ parameter describing a degree of magnetization is given by
\begin{eqnarray}
 \sigma = \frac{{\cal E}_{\rm EM}}{{\cal E}_{\rm MA}} .
\end{eqnarray}
The total specific energy and angular momentum are decomposed into
the electro-magnetic and matter (plasma) part and the corresponding subscripts
 are EM and MA, respectively. 
%%%
We add to note that $\eta$ and ${\cal E}$
reflect the amount of mass-loading/particle-injection into the jet 
\citep[e.g.,][]{Moscibrodzka11, Levinson11, Toma12, 
Hirotani16, Hirotani18, Chen18,  Levinson18, Parfrey19, Kisaka20}
although detailed studies on particle-injection
is beyond the scope of this paper.

The relativistic Alfv\'{e}n Mach number is defined as 
\begin{eqnarray}
M^{2} \equiv \frac{4\pi \rho u_{p}^{2}}{B_{p}^{2}}   .
\end{eqnarray}
The behavior of $M^{2}$ at the 
fast magnetosonic point is the key to understanding the
jet acceleration in the framework of MHD model.

\subsection{Relativistic Bernoulli equation}

Here we briefly review of relativistic Bernoulli equation.
The equation is also known as a poloidal wind equation.
Following  the framework of TT03, 
the normalized $r$ by the light-cylinder radius ($r_{\rm lc}$)
\footnote{The term $\hat{r}$ is denoted as $x$ in TT03 
since it focused on $r$-dependence of physical quantities.}
is introduced as 
\begin{eqnarray}
\hat{r}\equiv \frac{r} {r_{\rm lc}} =r\Omega_{F} , \quad
\Omega_{F}(\Psi)=const.  
\end{eqnarray}
In this work, we focus on the region where $\hat{r}>1$ holds.
%%%%%
%When discussing properties of MHD flows, 
%one may solve the Bernoulli equation for $M^{2}$ by assuming a curtain magnetic flux function 
%$\Psi(r,z)$ without solving the Grad-Shafranov equation, as a simplified step.
%In general, however, such solutions $M^{2}$ would not be
%trans-fast-magnetosonic.
%%%%%
If a poloidal velocity reaches the relativistic fast-magnetosonic wave speed
at a certain point, then the term $\partial M^{2}/\partial r$ 
may diverge at the point.
Such a flow solution is unphysical.
For a physical trans-fast magnetosonic flow solution,  
it is necessary to satisfy the critical condition there.
To remove this technical difficulty to find 
a special class of solutions of $M^{2}$ which satisfies this critical condition,
TT03 introduced a regular function of
\begin{eqnarray}
\xi(\hat{r})
\equiv \frac{E_{p}}{|B_{\phi}|}
\equiv \hat{r} \frac{B_{p}}{|B_{\phi}|}
\end{eqnarray}
where $E_{p}$ is the poloidal component of the electric field.
The $\xi$ is set as a smooth function of
 $\hat{r}$ including the fast magnetosonic point along  each given $\Psi=$constant surface.
It is worth stressing that $\hat{r}$ dependence on $\xi$, 
to be determined by the GS equation,
governs the magnetic field geometry and the corresponding velocity profile.

The Bernoulli equation is given by
\begin{eqnarray}
(1+u_{p}^{2})
(M^{2}+\hat{r}^{2}-1)^{2}&=& \nonumber \\
{\cal E}_{\rm co}^{2}(1-2M^{2}- \hat{r}^{2})
&+&
\left({\cal E}^{2}-
\frac{{\cal L}^{2}\Omega_{F}^{2}}{\hat{r}^{2}}\right)M^{4}
\end{eqnarray}
where the total specific energy measured in the co-rotation frame 
(${\cal E}_{\rm co}$) with the frame's rotation speed of $\Omega_{F}$ 
is defined as
\begin{eqnarray}\label{eq:e_0}
{\cal E}_{\rm co} &\equiv&  {\cal E} -{\cal L}\Omega_{F} .
\end{eqnarray}
By using the $\xi$, the Bernoulli equation reduces to 
the quadratic equation for $M^{2}$
\begin{eqnarray}
{\cal A} M^{4}-2 {\cal B} M^{2}+{\cal C} =0
\end{eqnarray}
where the coefficients  
${\cal A}$, 
${\cal B}$, and
${\cal C}$ are functions of 
$\Omega_{F}$, ${\cal E}$, $\xi^{2}$, and $\hat{r}^{2})$.
Readers can refer to
TT03 for details.
%%%%
Next, 
we consider Mach numbers at Alfven radius and fast magnetosonic radius. 
One can define the Alfven radius normalized by $r_{\rm lc}$ and
the Alfven Mach number at the Alfven radius as 
\begin{eqnarray}
M_{A} &\equiv & \frac{{\cal E} - {\cal L}\Omega_{F}}{{\cal E}} 
= 1- \hat{r}_{A}^{2}, \quad
\hat{r}_{A}\equiv \frac{{\cal L}\Omega_{F}}{{\cal E}} <1, 
\end{eqnarray}
which means the Alfven radius is within the light cylinder.
%
%We emphasize again that we do not have a singularity in $M^{2}$
%at $\hat{r}_{A}$ thanks to the introduction of the quantity $\xi(\hat{r})$.
%
Similarly, at the fast magnetosomic point ($\hat{r}=\hat{r}_{F}$),
\begin{eqnarray} 
M_{F}^{2}\equiv  \frac{(1-\xi^{2})\hat{r}^{2}+\xi^{2}}{\xi^{2}}
\end{eqnarray}
Thus, the behavior of the flow is controlled by the pitch angle of 
the magnetic field, which is reflected in $\xi$.
For instance, $\xi$ smaller than the critical value 
leads to $M^{2}= \infty$ at finite $r$.

\subsection{The approximated GS equation}

The approximated GS equation derived by TT03 (Eqs. (39) and (42) in TT03), 
which is valid for highly relativistic outflow 
of $1\le {\cal E}_{\rm co} \ll {\cal E}$,
is given by
\begin{align}
\frac{1}{\Psi_{r}}
\frac{\partial}{\partial r} 
\left[\frac{\xi^{2} \eta^{2}}
{1+   (\Psi_{z}/\Psi_{r})^{2}}
\left({\cal E}-\frac{{\cal L}\Omega_{F}}{\hat{r}^{2}}\right)
\right]
\simeq 
&& \nonumber \\ 
(1+\hat{M}^{2})\Omega_{F}^{2}
\frac{\partial}{\partial \Psi}
\left(
\frac{\eta^{2}}{\Omega_{F}^{2}\hat{M}^{4}}\right)   .
\end{align}
where $\partial \Psi(r,z)/\partial r = \Psi_{r}$ and
$\partial \Psi(r,z)/\partial z = \Psi_{z}$.
Since collimated jets in AGNs are discussed in the present work, 
the collimated geometry of the magnetic field is assumed  as follows:
\begin{eqnarray}
  \frac{\Psi_{z}}{\Psi_{r}} < \frac{1}{{\cal E}}\ll 1   .
\end{eqnarray}
Then one can obtain the general solution  of 
an analytical form for the approximated GS equation
(Eqs. (51) and (54) in TT03) described as
\begin{eqnarray}\label{eq:TT03-54}
\frac{2 (\hat{r}/{\cal E})^{2}(\hat{M}^{2}+1)}
{\hat{M}^{2}+2 (\hat{r}/{\cal E})^{2}} =
\ln \left[\frac{2 (\hat{r}/{\cal E})^{2}}{\hat{M}^{2}}+1\right]  .
\end{eqnarray}
From this, we can numerically obtain $\hat{M}(\hat{r})$.
By numerically solving the below
shown equation 
 (Eqs. (52) and (55) in TT03), one can obtain $\Psi(\hat{r},\hat{z})$ 
\begin{align}
\frac{\Psi}{\Psi_{0}}
 \left[\frac{1}{\hat{M} ^{2}(\hat{r}/{\cal E})} + \frac{1}{2 (\hat{r}/{\cal E})^{2}}\right]  
=
\frac{1}{\hat{M}^{2}(\theta_{0}\hat{z})} 
+ \frac{1}{2 (\theta_{0} \hat{z})^{2}} , 
\end{align}
where 
$\hat{z} \equiv z\Omega_{F}/\cal{E}$ and 
$\theta_{0}$ is the half opening angle of the jet
(see the next section).
It is well known that the geometry of magnetic field line is essential 
for jet acceleration and thus solving $\Psi$ is essential for discussing the velocity field.
Qualitatively,  magnetic field lines which bend towards the rotation axis
realize a location of $r_{F}$ at a finite distance from the central engine
\citep[e.g.,][]{Begelman94,Takahashi98}.

\subsection{Outer boundary wall condition}

The outer boundary wall condition would be given by
a parabolic streamline along the poloidal magnetic field lines. 
Since there are multiple normalization, it would be useful to explicitly write down the boundary condition here.
The outer boundary wall shape denoted as ($Z$, $R$)
in the cylindrical coordinate
and it satisfies the following relation:
\begin{eqnarray}\label{eq:boundary-wall}
\frac{Z}{r_{\rm lc}{\cal E}} \theta_{0} &=& \left(\frac{R}{r_{\rm lc}{\cal E}}\right)^{q} 
(1\leq q \leq 2),
\end{eqnarray}
where the 
$\theta_{0}$ is the half-opening angle of
the jet at the inlet boundary 
and the magnetic flux function on the boundary wall 
satisfies 
$\Psi(z=Z,r=R) = \Psi_{0}$.
Note that
the case of $q=1$ corresponds to 
a conical boundary wall shape.
%%
% To obtain a large specific energy E per one 
%particle, the rest-mass energy loading rate k per unit
%magnetic flux should also become small in proportion 
%to $\Psi$.
%%
In this work, we will give the value of $a$ 
with reference to the overall results of the detailed 
VLBI observations in \S 4.

\section{Basic properties of the velocity profile}

By solving these Bernoulli and GS equations, 
one can obtain a consistent $z$-dependence of $u_{p}$.
Before applying  TT03 model to the M87 jet,
here we overview the basic properties of  $u_{p}$. 
%%%
In \S~\ref{sec:Psi-dependence}, 
we show the $z$-profiles of $u_{p}$ for  
multi-streamlines with different $\Psi$.
In \S~\ref{sec:E-OmegaF-dependence}, 
we present
${\cal E} $ and $\Omega_{F}$ dependence of the $u_{p}$ profile
which will be important for comparisons of TT03 model with 
the observed $u_{p}$ of the M87 jet.

\subsection{$\Psi$ dependence}
\label{sec:Psi-dependence}

\begin{figure} 
\includegraphics
[width=18cm]
{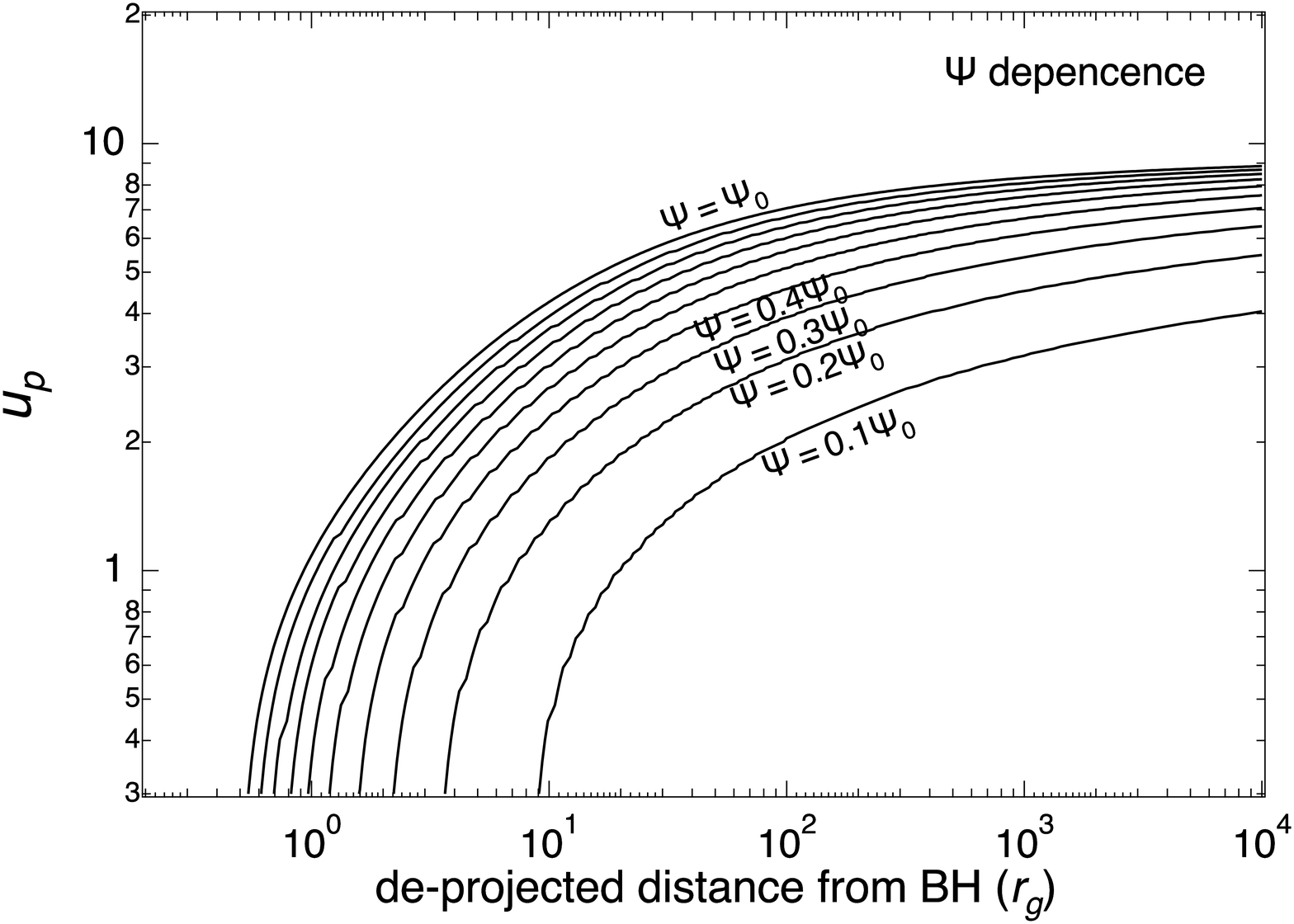}
\caption
{
We show the $\Psi$ dependence of $u_{p}$ by drawing the 
 $u_{p}$ profile with $\Psi=0.1\Psi_{0}, 0.2\Psi_{0},\dots, \Psi_{0}$.
 The fixed outer boundary wall has $q=1.3$. }
\label{fig:up_fiducial}
\end{figure}

Figure~\ref{fig:up_fiducial} shows the $z$-profile of $u_{p}$ for each streamline.
First, we briefly review the $r$-profile of  $u_{p}$ for
a given single $\Psi$=const. streamline.
One can define the square of the normalized relativistic Alfven Mach number
$\hat{M}^{2}(\Psi) \equiv \frac{M^{2}(\Psi)}{\hat{r}^{2}} = \frac{1}{\sigma(\Psi)}$.
It is also convenient to rewrite 
$\Gamma$ (equivalent to $u_{p}$) as
\begin{eqnarray}
\Gamma={\cal E} - \frac{u_{p}}{\xi \hat{M}^{2}}.
\end{eqnarray}
This shows $z$-dependence of jet acceleration 
by the energy conversion.
From Eq.~(\ref{eq:TT03-54}), 
in the inner zone $(1\ll \hat{r} \ll {\cal E})$, 
one can obtain
\begin{eqnarray} 
\hat{M}^{2}\approx \hat{r}/{\cal E}, \quad  
(1\ll \hat{r} \ll {\cal E})
\end{eqnarray}
This initial phase is identical 
to the linear acceleration phase indicated
by \citet[][]{Tchekhovskoy08}.
In the asymptotic far zone ($\hat{r} \gg {\cal E}$),  
the acceleration profile gets deviated 
from the linear acceleration and 
it becomes a logarithmic accelerated phase as 
\begin{eqnarray} 
 \hat{M}^{2}\approx 
 \ln\left(
 \frac{2 \hat{r}^{2}}{({\cal E}\hat{M})^{2}}\right), \quad
 (\hat{r} \gg {\cal E}),
\end{eqnarray}
the emergence of the logarithmic acceleration phase
after the linear acceleration phase 
is not only shown by TT03 
but also pointed out in
\citet[][]{Beskin98,Lyubarski09}.
The transition from linear to logarithmic acceleration is caused by plasma inertia.

As for  $\Psi$-dependence, faster $u_{p}$ is seen for larger $\Psi$ in Figure~\ref{fig:up_fiducial}.
 This behavior is explained by a differential bunching of $B_{p}$ in the jet.
 As already known in previous works of GRMHD simulations
 \citep[e.g.,][]{McKinney06, Nakamura18,Chatterjee19},
the energy conversion from ${\cal E}_{\rm EM}$ to ${\cal E}_{\rm MA}$ gets on 
at outer part of the jet flow with larger $\Psi$.
 Therefore, the  faster  $u_{p}$ is realized for larger $\Psi$.
 %%%%%%

%Again we emphasize that the advantage of TT03 is that the solution of TT03 satisfies 
%the trans-sonic flow and thus it can explicitly determine the location of fast point ($r_{F}$),
%while no other work can succeed to describe a trans-sonic flow in a semi-analytic form.

\subsection{${\cal E}$ and $\Omega_{F}$ dependence}
\label{sec:E-OmegaF-dependence}

\begin{figure} 
\includegraphics
[width=18cm]
{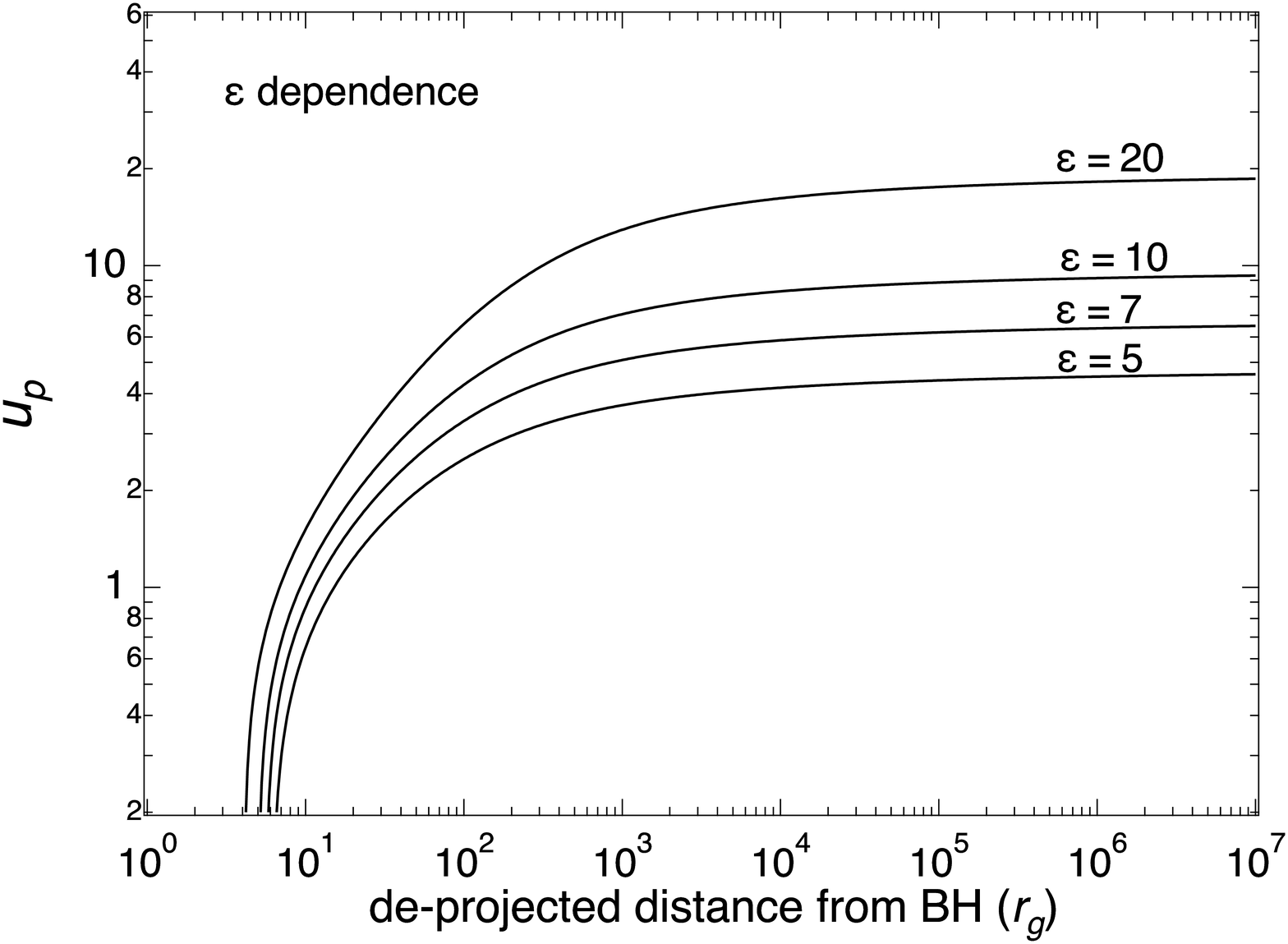}
\caption
{We show the ${\cal E}$ dependence of $u_{p}$. }
\label{fig:up_E}
\end{figure}

In Figure~\ref{fig:up_E}, we show the  ${\cal E}$-dependence of $u_{p}$.
Here we demonstrate the cases with ${\cal E}=$5, 7, 10 and 20. 
In order to demonstrate the ${\cal E}$-dependence , we select the flow with $\Psi=\Psi_{0}$
for each ${\cal E}$ case.
Since TT03 model does not include any energy dissipation processes,
it is clear that  the Lorentz factor asymptotically goes to the maximum value and it is given by
\begin{eqnarray}
{\cal E}=\Gamma_{\rm max} \quad (z\to \infty)    .
\end{eqnarray}
Since TT03 model describes an ideal magneto-transonic flow,
the complete energy conversion from ${\cal E}_{\rm EM}$ to ${\cal E}_{\rm MA}$
realizes at infinity.
%%%
%Since the $z$-profile of  the Alfv\`{e}n Mach 
%number follows
%$\hat{M}^{2}\approx \ln(2 \hat{r}^{2}/({\cal E}\hat{M})^{2})$,
%The larger ${\cal E}$ becomes, 
%the larger the distance $z$ becomes for reaching to 
%the terminal Lorentz factor.

\begin{figure} 
\includegraphics
[width=18cm]
{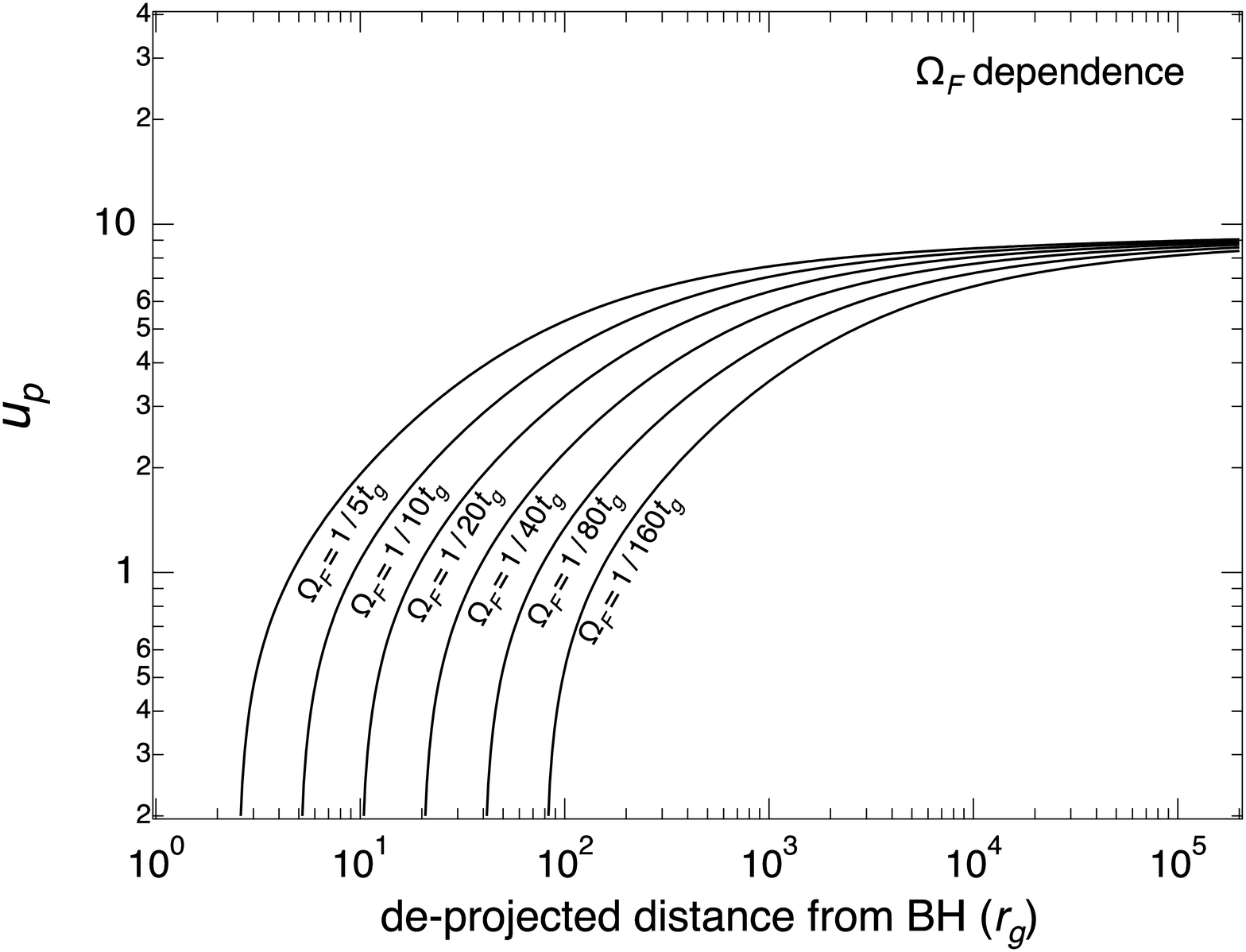}
\caption
{We show the $\Omega$ dependence of $u_{p}$. }
\label{fig:up_OmegaF}
\end{figure}

In Figure~\ref{fig:up_OmegaF}, we show the  $\Omega_{F}$-dependence of $u_{p}$.
Same as Figure~\ref{fig:up_E}, we select the flow 
with $\Psi=\Psi_{0}$ for each $\Omega_{F}$ case to present the $\Omega_{F}$-dependence.
On the contrary to the case of varying ${\cal E}$,  
$\Omega_{F}$ does not 
alter the profile of $u_{p}$ itself.
%%%
As already shown, $\Omega_{F}$ is governed by
the light cylinder radius $r_{\rm lc}$ and it is given by
$\Omega_{F}= 1/r_{\rm lc}$.
%
%%%%%
%For a  given assumed boundary condition, the value $X_{F}$
%can be determined such as  $X_{F}=1/2$ for 
%matching Michael solution at infinity.
%%%%%
Slower rotation of $ \Omega_{F}^{-1}$ 
leads to more distant starting point of the jet acceleration from the central BH.

\subsection{Location of intersection
between Boundary-wall and light-cylinder}

In this work, a location of intersection between 
the boundary-wall ($\Psi=\Psi_{0}=1$ surface)
and the light-cylinder is important. Hereafter,
we denote the location as ($R_{0}$, $Z_{0}$).
By inserting $R_{0}=r_{\rm lc}$ at Eq.~(\ref{eq:boundary-wall}),
the location of $Z_{0}$, from which 
the jet acceleration starts  (see Figure 1), 
is obtained as follows:
\begin{eqnarray}
Z_{0}=100~r_{g}
\left(\frac{\theta_{0}}{0.1}\right)^{-1}
\left(\frac{r_{\rm lc}}{10r_{g}}\right)
{\cal E}^{1-q}   ,
\end{eqnarray}
where it will turn out to be ${\cal E}^{1-q} \approx 0.2 - 0.5$ 
from observational properties of the M87 jet shown in the next section.
We note that the geometrical factor $\theta_{0}$ also affects the location of $Z_{0}$.

\section{Application to the M87 jet}\label{sec:Application}

Here we apply TT03 model to the  $u_{p}$ measured in the M87 jet.
%%
%Note that the values of $\eta$ and ${\cal L}$ 
%are not explicitly treated in the present work.
%%
 The two parameters to be determined are  ${\cal E}$, and $\Omega_{F}$.
While ${\cal E}$ is easily constrained from 
the maximum speed  of the M87 jet around HST-1 region, 
$\Omega_{F}$ has been poorly constrained by any observational data so far.

\subsection{Maximum Lorentz factor}
%%%
As mentioned, in the asymptotic zone ${\cal E}$ satisfies ${\cal E}\approx \Gamma_{\rm max}$.
Regarding the  maximum Lorentz factor, 
it is chosen to match with the HST-1 component 
at the $4\times 10^{5}~r_{s}$ 
\citep{Biretta99, Giroletti12}
and we set
\begin{eqnarray} 
{\cal E} = 10  .
\end{eqnarray}
To clear up the essential discussion in this work, 
hereafter we fix the value ${\cal E}=10$ for simplicity, 
which never affects the main result of this work.

The jet half-opening angle is directly
constrained by VLBI observations
\citep{Junor99, Hada16}.
Here, we set 
\begin{align} \label{eq:theta_0}
\theta_{0}
\approx 0.34~\tan^{-1}\left[
\sin\left(\frac{\theta_{\rm view}}{17^{\circ}}\right)
\tan\left(\frac{\theta_{0, \rm obs}}{50^{\circ}} \right)\right]~{\rm rad}
\end{align}
in our subsequent numerical calculations.
The chosen value of the observed
$\theta_{0, obs}$ is adopted from the result 
of the high dynamic range VLBA+GBT obseravation
at 86~GHz, which indicate the $\sim 100^{\circ}$ 
of the full-opening angle at the jet base
\citep{Hada16}.
The viewing angle of the M87 jet 
$\theta_{\rm view}$ has uncertainty and here
we adopt the normalization of $17^{\circ}$ 
based on \citep{EHT19_5}.

\subsection{Boundary-wall shape}

In the present work,
we identify the jet width profile measurement conducted by 
\citet[][]{Asada12, Hada13} as the boundary-wall shape.
It means that the observed M87 jet is
identical to the funnel region, in which the ordered magnetic field collimates and accelerates the plasma jet.
%%%
It is, however, difficult to know exactly which magnetic field lines among the ordered fields
are identical to the observationally measured jet profile.
If no dissipation occurs at the boundary between the jet and surrounding matter
and the funnel region is filled with radio emitting non-thermal electrons,
then 
the outermost magnetic field lines on $\Psi=\Psi_{0}$ surface
that thread the black hole are basically identical to the jet width profile measured by VLBI described above.
In more realistic cases, however, 
the boundary region between a jet and a surrounding matter may become dissipative by reflecting 
the details of physical conditions at the boundary layer \citep[e.g.,][]{Levinson16, Chatterjee19}.
It is also uncertain about where and how nonthermal electrons are produced and cooled down in the jet.
Thus, model predicted images generally depend on assumptions in treatments of 
non-thermal electrons
\citep[e.g.,][]{Dexter12, Takahashi18, EHT19_5}.
Therefore, taking those uncertainties into account,
we include all the allowed range of $q$ 
obtained by \citet{Asada12, Hada13} is as follows:
\begin{eqnarray}
1.3 \lesssim  q  \lesssim    1.7 .
\end{eqnarray}
%
%Closer to the black hole, the measured radial profile suggests a possible change in the jet
%collimation shape from the outer parabolic one, where the jet shape tends to become more radially oriented. 
%This result has been indeed confirmed by a high dynamic range observation 
%of VLBA+Green Bank Telescope at 86~GHz (Hada et al. 2016).
%%
In addition, TT03 model can describe the magnetic field bending around 
the characteristic distance $z_{\rm brk}$ where ${\cal E}_{\rm EM}\sim {\cal E}_{\rm MA}$ holds.
It is obtained by the condition of $ \hat{r}\approx{\cal E}$ at $R=R_{\rm brk}$ 
and given by
$ z_{\rm brk} = 1000~r_{g}
\left(\frac{{\cal E}}{10}\right)
\left(\frac{\theta_{0}}{0.1}\right)^{-1} 
\left(\frac{r_{\rm lc}}{10~r_{g}}\right)$
in the deprojected distance (see eq. (56) in TT03).
%%
%The possible break feature in the jet-width profile
%reported in Hada et al. (2013) at $z_{\rm brk}  \approx 100~R_{s}$.
%Although there are a factor of several deviation in its location,
%we can say that it shows a good agreement as a first order approximation.
%%%
%Since TT03 model is structured within the framework of special relativity
%and we have the deviation in $u_{p}$ between the model and observations,
%we do not over-explain possible reasons of this deviation. 

\subsection{Velocity profile}

Here, we fit the observed  $u_{p}$ 
data with the one predicted by TT03 model.
At first, we explain how to do the fitting.
As already mentioned, 
we fix ${\cal E}=10$ throughout this work. 
Therefore, the remaining model parameters
to be adjusted are $\Omega_{F}$ 
(equivalent to $r_{\rm lc}$) and $\theta_{0}$.
%%%
To properly search for the best fit 
$\Omega_{F}$ and $\theta_{0}$
taking the uncertainty into account, 
we impose the following condition.
\begin{enumerate}
\item 
From the currently measured range of 
$u_{p}$ \citep[][and references therein]{Park19}, 
we set the allowed range of $Z_{0}$ as 
$2\times 10^{2}~r_{g}\lesssim 
Z_{0} \lesssim 4\times 10^{2}~r_{g}$ 
in this work.

\item
Based on VLBI observations, 
we set the allowed range of $\theta_{0}$ 
as $0.1 \le \theta_{0} \le 0.34$ 
where the upper bound is given by Eq.~(\ref{eq:theta_0}) 
while the lower bound is assumed as
$\theta_{0}=1/{\cal E}=0.1$.

\item
The light-cylinder radius should be smaller 
than the jet radius in the jet acceleration region
(i.e., $z \gtrsim Z_{0}$)
within the framework of TT03 model.
The jet width $R_{0}$ at $z \approx Z_{0}$ is 
measured as $R_{0}\approx 100~r_{g}$
\citep[][]{Hada13}.
\footnote{The jet width (Full Width Half Maximum) at $z\sim Z_{0}$
presented in \citet[][]{Hada13} can be approximated as 
$2R_{0}\approx 100~r_{\rm s}$ 
where $r_{\rm s}=2r_{g}$ is the Schwarzschild radius.}

\item
As shown in Figure~\ref{fig:up_OmegaF}, 
$\Omega_{F}$ tightly links to $Z_{0}$.
We will search for a best fit $\Omega_{F}$ 
so that the model predicted $u_{p}$ of 
the outer edge of the jet flow ($\Psi=\Psi_{0}$) 
does not largely exceed the observed data.

\item
Taking various uncertainties into account, 
here we will perform the fitting for 
both $q=1.3$ and $q=1.7$ boundary-wall 
conditions and we will determine the allowed range 
of $\Omega_{F}$ in between those two best-fit values.
\end{enumerate}

%\subsubsection{The case of q=1.3$}

\begin{figure} 
\includegraphics
[width=18cm]
{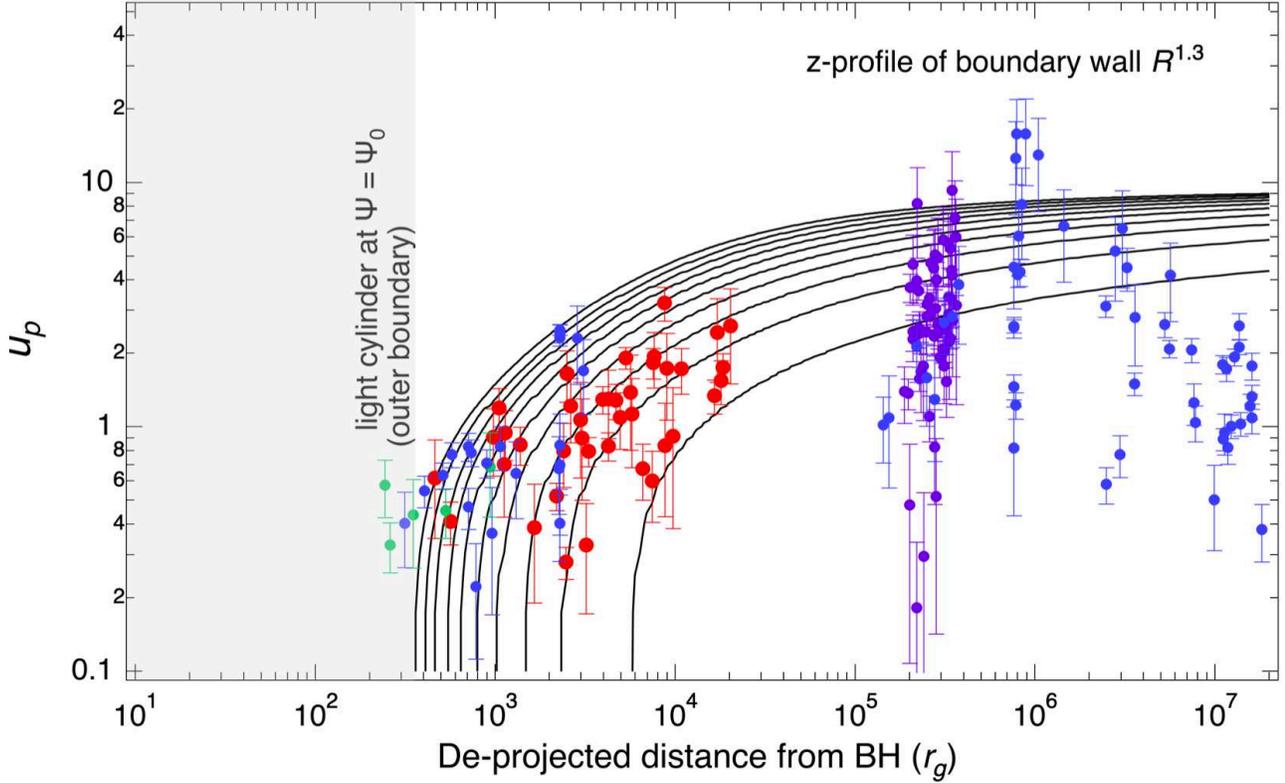}
\caption
{
The comparison of the model predicted  $u_{p}$ and
and the VLBI measured  $u_{p}$ as a function of 
deprojected distance $z$ from the black hole in unit of $r_{g}$. 
The observation data points are adopted from \citep[][]{Park19}.
The case in which the parameter $q=1.3$ is chosen for the boundary wall shape 
is presented here.
%%
%We also include the results obtained by GRMHD simulations shown with the purple open upward triangles 
%(McKinney 2006; Penna et al. 2013; Nakamura et al. 2018). 
%%
Whereas TT03 model well reproduces the overall logarithmic acceleration
of $u_{p}$,  
some offset remains at the inner region $z\lesssim 500~r_{g}$.
We also note that it is natural for the observed data and the model to have a gap
below $10^{6}~r_{g}$ since the region is beyond the domain of application of the model, where various dissipation processes are supposed to occur.}
\label{fig:up_best13}
\end{figure}

In Figure~\ref{fig:up_best13}, we show the best fit profile of the $u_{\rm p}$ for the case of the boundary-wall with $q=1.3$.
%%%%%%%%%%%
Same as Figure~\ref{fig:up_fiducial}, 
we have plotted the multiple flow paths along 
with $\Psi=0.1\Psi_{0}, 0.2\Psi_{0},\dots, \Psi_{0}$.
%%%%
As shown in Figure~\ref{fig:overall},
the jet flow is not heterogeneous but described as multiple laminar flow paths along multiple magnetic surfaces. 
%%%%%
The jet plasma is not essentially accelerated inside the light cylinder. 
That is because the plasma is co-rotating with the magnetic field lines inside the light cylinder. 
%%%%%%
When the plasma exceeds the light cylinder radius, it cannot co-rotate with the magnetic field lines anymore and flows outward, and is accelerated in the poloidal direction. 
%%%%%%%%%%%
The KaVA observational data shown in \citet{Park19} indicates that the M87 jet logarithmically accelerates up to the HST1 scale.
Hence the linear acceleration \citep[e.g.,][]{Tchekhovskoy08} 
is not able to explain the observed $u_{\rm p}$ profile.
%%%
On the contrary,
TT03 model predicts the logarithmic acceleration, which naturally
agrees with the observed logarithmic $u_{\rm p}$ profile.
%%
%Since $\hat{r}\gg {\cal E}=10$ holds for the VLBI observed region
% (i.e.,  $\hat{r} > 10^{2}$), 
%%
As pointed out by \cite{Park19},
the observed trend of the jet acceleration in M87 is slower 
than those indicated in GRMHD simulation in the literature
and does not match each other.
From Figure~\ref{fig:up_best13}, one can find that 
our model can overcome this problem and 
explain with this observed velocity profile above  $10^{3}~r_{g}$ scale.
The reason is the best fit parameter
$r_{\rm lc}\approx 70~r_{g}$ is 
larger than a typical one in GRMHD simulations.
For instance, GRMHD simulation of the M87 jet 
in \citet[][]{McKinney06}
obtained $r_{\rm lc} \approx 10~r_{g}$.
%%%
We thus find that the larger $r_{\rm lc}$  
shifts the starting point 
of the jet acceleration, i.e., $Z_{0}$ and 
it can explain the observed $u_{p}$.
The obtained distance is $Z_{0} \approx  3\times 10^{2}~r_{g}$.
%%%
We add to note that the condition 3 holds only when $\theta_{0}\approx 0.1$. Hence we use this value 
although this is a factor of $\sim 3$ 
smaller than the  $\theta_{0}$ indicated by 86GHz observation.
An investigation of this mismatch is
beyond the scope of this work since TT03 model cannot 
discuss anything inside the light cylinder.
%%%

In Figure~\ref{fig:up_best13},
one can see that the innermost data points within 
$30~r_{g}$ do not match the model prediction. 
But it is not fatal. Although the detailed investigation is beyond the scope of this paper, 
this mismatch may suggest the need for effects not incorporated in TT03 model. 
In \citet[][]{takahashi21}, we discussed that differences in the angular momentum values of the plasma are one possibility to mitigate the discrepancy.

%\subsubsection{The case of $p=1.7$}

\begin{figure} 
\includegraphics
[width=18cm]
{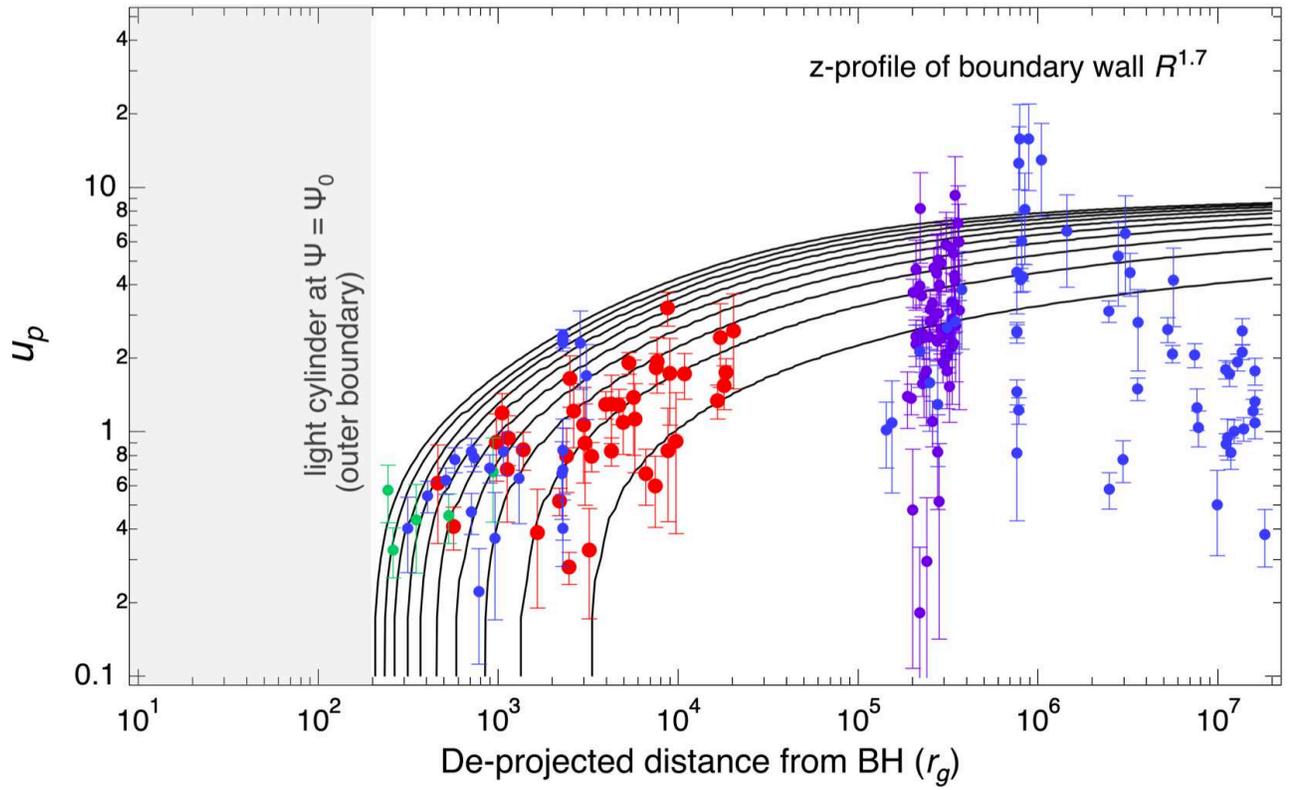}
\caption
{Same as Figure~\ref{fig:up_fiducial} but with $q=1.7$. }
\label{fig:up_best17}
\end{figure}

In Figure~\ref{fig:up_best17}, we show the best fit profile of 
the $u_{\rm p}$ for the case of the boundary-wall with $q=1.7$.
The overall behavior of the the model predicted $u_{p}$ profile
is similar with the case with  $q=1.3$.
The difference between $q=1.3$ and $q=1.7$ cases
is that the case with  $q=1.7$ have a smaller (more gradual) slope of acceleration
than the case with $q=1.3$.
The best fit value adopted in Figure~\ref{fig:up_best17} is 
$r_{\rm lc}\approx 100~r_{g}$
and correspondingly
we have 
$Z_{0}\approx 2\times 10^{2}~r_{g}$ in Figure~\ref{fig:up_best17}.

Finally, by setting the result for the case of $q=1.3$ 
as the upper limit of $\Omega_{F}$
and setting the result with $q=1.7$ 
as the lower limit of  $\Omega_{F}$,
we obtain the allowed range of $\Omega_{F}$ as follows:
\begin{eqnarray}\label{eq:omega_f}
\frac{c}{100~r_{g}}\lesssim \Omega_{F} \lesssim \frac{c}{70~r_{g}}  .
\end{eqnarray}
Thus, we find that 
the slower $\Omega_{F}$ compared to typical values in GRMHD simulations 
mitigates the velocity mismatch problem 
in the M87 jet pointed out in \citet{Park19}.

\subsection{$\sigma$ profile}

\begin{figure} 
\includegraphics
[width=18cm]
{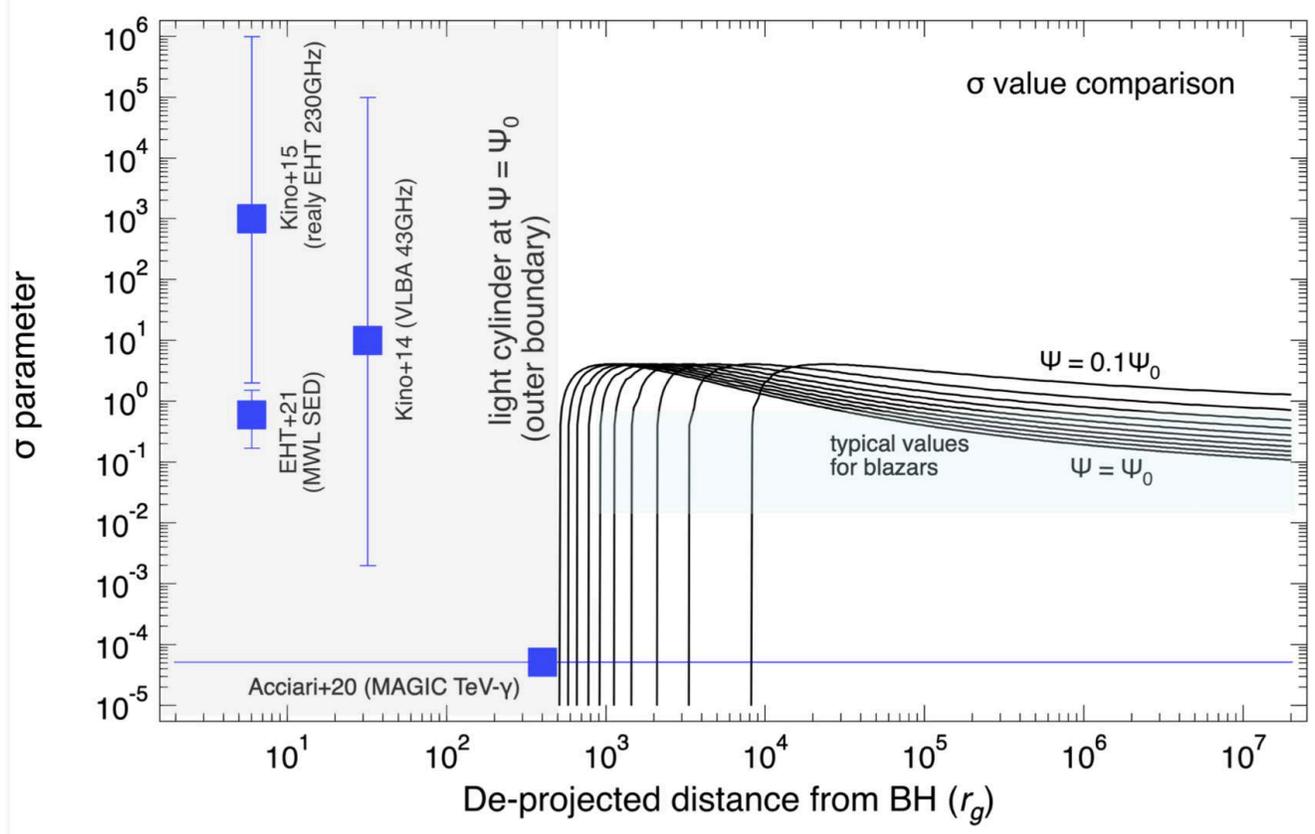}
\caption
{The comparison of the obtained $\sigma$ profile for each magnetic field lines and 
the estimations of the magnetization degrees constrained in previous literature for the M87 jet.
For comparison, a region indicating typical values for the location and the magnetization degree of blazars is also shown by a light blue filled square.
Unfortunately, the estimates made by \citet[][]{Kino14, Kino15}
are in a region beyond the applicability of TT03 model (i.e., inside the light cylinder), 
which means that they cannot be directly compared with the 
obtained $\sigma$ curves presented here.
The result reported by \citet[][]{Magic20} 
indicates apparently much smaller 
magnetization degree than the values
found in other reports.
}
\label{fig:sigma_best13}
\end{figure}

In Figure~\ref{fig:sigma_best13},  we show the corresponding 
$\sigma$ profile with $q=1.3$
together with the $\sigma$ values obtained in literatures.  
\footnote{
The part of sharply rising $\sigma$  at small $z$ should be neglected 
since this is the unphysical branch of the solution, 
which also appeared in 
\citet{Takahashi98}.}
%%
%Since $\Gamma\approx 1$ holds at the inlet boundary of the jet,
%one can obtain  $\sigma ={\cal E}/c^{2}$.
%
In general, there is a limitation for constraining a magnetization degree 
at a jet base from spectral energy distribution  (SED) fitting of 
multi-wavelengths (MWL) data since collected flux data 
do not share a common single emission region due to different angular resolution 
of various telescopes.
%Nevertheless, we include the recent report of MWL  SED fitting 
%by\citet[][]{Magic20} as a best effort basis estimate.
%%%
To overcome the limitation of MWL SED fitting, 
\citet[][]{Kino14} and \citet[][]{Kino15} 
explore the energetics at the M87 jet base
based on VLBI data alone together with the well-established 
process of synchrotron self-absorption (SSA).
%%%
%We obtained that 
%$1\times 10^{-5} \lesssim U_{B}/U_{\pm}\lesssim 6\times 10^{3}$ 
%within the radio core seen with VLBA at 43~GHz, and 
%$5 \lesssim U_{B}/U_{\pm}\lesssim 1 \times 10^{6}$ 
%within the putative SSA-thick region in EHT emission region at 230~GHz
%where $U_{B}=|\vec{B}|^{2}/8\pi$ and $U_{\pm}$  is 
%the energy density of total magnetic field and 
%that of non-thermal electrons and positrons, respectively.
%%%
In Figure~\ref{fig:sigma_best13},  we include these values in the
literatures by setting 
$\sigma={\cal E}_{\rm EM}/{\cal E}_{\rm MA} \approx U_{B}/U_{\pm}$.
%%%%%
%It corresponds to the case in which  the energy 
%density of non-thermal electrons (and positrons) is comparable to that of the 
%bulk kinetic energy for the cold plasma flow dealt in TT03.
%%%%%
Unfortunately, sub-mm radio-emitting 40~$\mu$as region in 
\citep[][]{Kino14, Kino15, EHT_MWL21} is within the light cylinder, 
which is not described by TT03.
Therefore, it is not possible to directly compare the obtained 
$\sigma$ profile with the constrained $\sigma$ 
those previous works.
At least, what one can conservatively say is that
a high value of $\sigma$ inside the light cylinder does not contradict to 
the overall picture of the magnetic acceleration of the jet.
We also plot the resultant $\sigma$ by MWL SED fitting 
\citep[][]{Magic20} plotted in  Figure~\ref{fig:sigma_best13}
 shows extremely low magnetization degree to explain the
 observed $\gamma$-ray emission.
 To explain this, an extremely efficient conversion process
 from Poynting flux into kinetic one is required.
 %
%***
%At the same time, we should bear in mind that 
%a possibility that the indicated extremely low magnetization 
%by Acciari et al. (2020) is an artifact caused by the multi-wavelengths 
%SED which possibly is a superposition of emissions from different regions.
%*** 

\section{Discussions}

\subsection{Observational evidence of the boundary-wall}

First, we begin with a recent observational 
support for the existence of the global wind component in M87, 
which plays a role of the outer-boundary wall 
that confines the jet.
%%%
The need for such an outer wall has been generally
suggested in theoretical studies \citep[e.g.,][]{Nitta97}.
A wind component is naturally considered to play 
the role of an outer boundary-wall.
Observataionally, a parabolic shape is required 
as the boundary-wall.
%%%
The existence of the wind component surrounding the M87 jet
has been indeed discovered by \citet[][]{Park19a}
using eight VLBA data sets, one at 8 GHz, four at 5 GHz, and three at 2 GHz. 
Faraday rotation measures (RMs) measured across the bandwidth of each data set
were obtained and the authors found that the magnitude of RM systematically
decreases with increasing distance from 5000 to 200,000 Schwarzschild radii. 
The data, showing predominantly negative RM signs without significant difference 
of the RMs on the northern and southern jet edges, suggest that the spatial extent 
of the Faraday screen is much larger than the jet. 
\citet[][]{Park19a} find that the decrease of RM along the M87 jet axis
 is described well by a gas density profile that is inversely proportional to $z$.
This observational data support the collimation of the M87 jet by the surrounding winds. 
\footnote{
Detailed structures of wind components may be different for each object and it is still under debate \citep[e.g.,][]{lisakov21,Okino22}.}

\subsection{Comparison with force-free jet model}\label{sec:FFmodel}

The suggested value of $\Omega_{F}$ in Eq.~(\ref{eq:omega_f})
is somewhat slower than the typically claimed 
$\Omega_{F}\approx \Omega_{H}/2$ in force-free jet models.   
Hence, it is worth discussing the possible 
origin of the difference.
%%%%%
%The differences in previous force-free jet models may 
%be understood.
%%%%
For monopole magnetic field,
the exact solution of $\Omega_{F} = \Omega_{H}/2$
is obtained by equating 
Michel's monopole magnetic field solution 
$B_{\phi}=B^{r} \Omega_{F} \sin \theta$
obtained by the outer-infinity boundary condition
\citep[][]{michel73}
with the Znajek's inner boundary condition
\citep[][]{Znajek77} on the event horizon
$B_{\phi}=B^{r}(\Omega_{F}-\Omega_{H}) \sin \theta$
\citep[see details for][]{komissarov04,Beskin10}.
%%%
Generally, $\Omega_{F}$ depends 
on magnetic field configuration and
current density distribution in the magnetosphere
\citep[e.g.,][for review]{Beskin10}. 
%%%
%and BZ77 also derive the analytic form of  $\Omega_{F}$ for %parabolic magnetic field case, 
%(see also Eq.~(3.101) in Beskin for review). 
%In the case of BZ77, an (artificial) 
%current sheet is required on the equatorial plane.
%%%
%and they obtain the arrowed range as $0.265\lesssim \Omega_{F}/\Omega_{H}\le 0.5$
%%%%
%Other cases in which the longitudinal current is iteratively solved have been explored.
%Beskin et al. (1992) studied the case where the central black hole is surrounded by
%the well-conducting disk.
%%%%%
\citet{nathanail14} 
studied $\Omega_{F}/\Omega_{H}$ for
the magnetic field lines extend from the Kerr black hole
with a thin disk (current sheet) that sources toroidal current.
Their results showed 
$\Omega_F/\Omega_H \gtrsim 0.2$ (see their Figure 2). 
In \citet{nathanail14}, their numerical procedure for solving $\Omega_{F}$ and longitudinal current 
works only for the case when the light cylinder is not too far away due to the numerical
box size. 
That was probably why they chose the initial condition as  $\Omega_{F}=\Omega_{H}/2$.  
\citet{Ogihara21} studied 
the case where the density floor problem is alleviated
by solving the transverse force balance between
the field lines at the separation surface 
and they suggests $\Omega_{F}\approx (0.35-0.5)\Omega_{H}$.  
%%%%%%
%The important point to bear in mind is that the outer infinite boundary condition 
%$B^{\phi}=E_{\theta}$ (1D steady state force-free solution) is adopted.
%%%%%%%
%A global  model of geometry of BH magnetosphere was constructed by BZ77 
%within  force-free (FF) limit with the  assumption of slowly rotating BH.
%The FF magnetosphere generate poloidal current that forms a circuit
%and BH's rotation energy can be extracted in a form of Poynting flux.
%%%
%In the framework of TT03, the light cylinder is set as the boundary and
%an accuracy of the solutions becomes lower near the boundary surface.
%%
%Since the emission region is dominated by the energy density of magnetic fields (Kino et al. 2015),
%obtaining the force-free (FF) BH magnetosphere is one of the essential tasks to get the TT03 framework better.
%%%
%By performing the time-dependent GRMHD simulation with
%the magnetically dominated split monopole magnetosphere,
%Komissarov (2014) confirmed that
%that the numerical solution indeed evolves towards a stable steady-state solution which is very close to 
%the corresponding FF solution found by Blandford \& Znajek.
%%%
\citet{Thoelecke17, Thoelecke19} also investigated 
steady force-free magnetic field configuration 
without placing a thin current sheet 
on the equatorial plane nor imposing 
outer-infinity boundary condition.
They found that the resultant configuration
are classified into the following three cases:
(i) conical jet and wind structure 
appears when the BH-spin is slow or $\Omega_{F} \sim \Omega_{H}/2$, 
(ii) conical jet and wind structure realizes when 
the fast BH-spin with slow $\Omega_{F}$.
(iii) equatorial wind structure realizes when both $\Omega_{H}$ and $\Omega_{F}$ are high.
Our suggestion of $\Omega_{F}$ agrees with the case (ii).
Therefore, the relatively slow $\Omega_{F}$ in Eq.~(\ref{eq:omega_f})
suggested in this work could indicate that the outer boundary condition is
different from Michel's monopole solution 
or the absence of a thin current-sheet on 
the equatorial plane in M87.

\subsection{Comparison with GRMHD simulations}

\subsubsection{Comparison between disk-jets model and wall-jets model}

%It is worth checking influences of artificially 
%placed ideal boundary walls.
Recently \citet{Chatterjee19}  made a comparison between the
wall-jets model and the disk-jets model.
\citet{Chatterjee19}  refer to simulations where
a jet is surrounded by an idealized
perfectly conducting external boundary-wall
as wall-jets model, 
while we call it disk-jets model without 
such an artificial ideal boundary-wall.
%%%%%%%%%%%%%%
%the idealized  MHD flow 
%confined by the boundary-wall of external medium and 
%full GRMHD jet simulations and to know 
%%%%%%%%%%%%%
%A pioneering work of \citet{McKinney06} 
%performed a long-distance 
%GRMHD simulation of the disk-jet model and clarified 
%its basic properties.
%%%%%%%%%%%%%
\citet{Chatterjee19} showed
that $\Gamma$, $\sigma$, and ${\cal E}$  agree well
between the disk-jet and the wall-jet.
It means that the wall-jets model well 
capture most of the time-averaged steady-state 
properties of the disk-jet model  
with the same shape in the absence of instabilities. 
%%%%
Although it does not affect the main result of this work,
there is an interesting difference between 
the wall-jets model and the disk-jets model.
For the disk-jets model, the presence of
a pressure imbalance between the jet and the accretion disk-wind
gives rise to oscillations in the jet shape.  
It causes a difference
in the value of enthalpy between the two setups.
For the disk-jet model, 
the enthalpy increases substantially at 200~$r_{g}$
due to the onset of the pinch instabilities that convert the poloidal
field energy into enthalpy and this is the main difference between these two different setups.

%%%%%
%Although  \citet{Chatterjee19} 
%made a detailed investigation about  
%pinch instability along with the $z$-direction,  
%a thorough systematic study on $r$-dependence
%(one may say $\theta$-dependence in spherical coordinate) 
%of the physical quantities are missing.

\subsubsection{Validity of constant $\Omega_{F}$}

Next,
we discuss the validity of the assumption of a constant  $\Omega_{F}$.
\citet{Komissarov07}
conducted special relativistic MHD
jet simulations in which
the jet is confined by a rigid boundary wall.
%%
%whose boundary-wall and inlet-boundary conditions is similar with that adopted in TT03 model.
%%
As for the inlet boundary condition, \citet{Komissarov07}
explored the 
two cases for $\Omega_{F}$, i.e., solid-body rotation and differential rotation.
The solid-body rotation law (i.e.,  $\Omega_{F}=$const.) 
would provide a good description of magnetic 
fields that thread the event horizon of a central BH, 
while the differential rotation law is more suitable when
the magnetic fields anchor to the accretion disk.
These two cases reproduce the different distribution in $u_{p}$.
The constant $\Omega_{F}$ case 
\citep[C2 model in][]{Komissarov07})
shows faster velocity field for larger $\theta$ ( i.e., near the outer-boundary wall), 
while slower velocity realizes for smaller $\theta$ (i.e., near the jet axis).  
On the other hand,
the differential rotation of  $\Omega_{F}$ 
\citep[C1 model in][]{Komissarov07})
realize the inverse situation, i.e., slower velocity near the  outer-boundary wall, 
while the flow has faster velocity near the jet axis.
It is clear that the resultant $u_{p}$ of TT03 model is in good agreement with the 
model of C2 in  \citet{Komissarov07}.
\citep[][]{Nakamura18}
also made a detailed study on $\theta$-dependence 
of $u_{p}$ in a more realistic case by performing GRMHD simulation.
They also found the same result with  \citet{Komissarov07}, 
i.e., 
shows faster $u_{p}$ for larger $\theta$
( i.e., faster sheath)
and slower  $u_{p}$ for smaller $\theta$ ( i.e., slow spine).
Thus, the validity of the assumption of a constant  $\Omega_{F}$
is reasonably supported by GRMHD simulations.

\subsection{Possible origin of the slow $\Omega_{F}$}

Here, we discuss possible origin of the slow $\Omega_{F}$.
As briefly discussed in \S~\ref{sec:FFmodel}, 
choice of boundary conditions would generally
affect the value of $\Omega_{F}$.
%- 定常MHDモデルで記述できるとして、Omega_F << Omega_Hはホライズン付近に
%生じるギャップで電位降下していることを示しているのかもしれない 

One possibility that we would like to bring up first 
is an injection of plasma  
generated at an unscreened strong electric field regions 
(so-called "vacuum gaps") along magnetic field line 
may form close to the horizon in under-dense 
black hole environments in which the force-free approximation breaks down
\citep[e.g.,][]{hirotani98,Chen18, Hirotani18,katsoulakos20}.
%%%
\footnote{ 
Regarding the ergosphere region,
\citep[][]{toma14} pointed out that $\Omega_{F}$ 
is deduced for the field lines threading the equatorial 
plane in the ergosphere
by considering open magnetic field lines penetrating the ergosphere that keeps driving the poloidal currents and generating the electromotive force and the outward Poynting flux.}
%%%%
Charged seed electrons, injected by 
e.g. pair-creation processes
(in an inner accretion flow) into these regions, 
are then quickly accelerated along the fields 
to high energies 
and can trigger an electromagnetic pair cascade and that eventually ensures 
a charge supply high enough to establish the formation of jet like features. 
%%%
Within the framework of steady MHD flows,
the slower $\Omega_{F}$  would  require a larger total angular momentum
${\cal L}$ when $\Omega_{F}{\cal L}\sim {\cal E}$ holds
and this is indeed shown by the recent work of \citep[][]{takahashi21}.
Therefore, we can say that
how to achieve such ${\cal L}$ at the plasma injection point 
will be a key question to be solved in future work.

A recent study of \citet{Levinson17} explored steady gap solutions 
around Kerr BH and they showed that such solutions are allowed 
only under restrictive conditions 
and then they conclude that magnetospheric gaps are intermittent. 
Such intermittency could affect floor conditions 
in GRMHD simulations and it could help slowing 
down the $\Omega_{F}$.
However, it should be fair to note that the gap model 
do not successfully reproduce sufficient amount of $e^{\pm}$ pairs to explain a required total jet power \citep[][]{Kisaka20}.
%%%
In contrast, 
drizzle pair production cascade model predict that 
smooth background of MeV photons produced by a hot accretion flow 
that interact with each other and  $e^{\pm}$ pairs are produced \citep[][]{Moscibrodzka11}.
Recently, \citet[][]{wong21drizzle} revisited the drizzle model  using radiative GRMHD.
They found that the drizzle pair production process produces a background pair above 
the Goldreich–Julian (GJ) 
density in M87-like SANE model that may make it difficult to open the gap.
To obtain a consistent picture, combined study of the gap and the drizzle models would be of great importance in future work.

Second possibility is due to the difference in an 
outer torus 
%%%%%%%%
\footnote{
The term ``torus" used here is identical to that widely used in studies using GRMHD numerical simulations. They are sometimes called SANE torus or MAD torus 
\citep[e.g.,][]{Murchikova22}. 
These tori play a role in supplying mass and magnetic fields onto the central BH. These tori set up in the GRMHD simulation have not yet been directly observed, little is known about their 
observational properties and
their relations to molecular tori.}
%%%%%%%%%%
that feeds the magnetic fields into the funnel region.
The magnetosphere in the funnel region is built up via the accreted magnetic fields from the geometrically-thick hydrostatic outer torus put in GRMHD simulations.
Therefore, the property of the outer torus would 
affect the magnetic field in the funnel region. 
%%%
%The B field configuration, or the shape of the boundary wall, and thus
%that may affect the value of $\Omega_{F}$.
%%
The torus with the constant angular momentum was considered  \citet[][]{Fishbone76,Kozlowski78}.
and it is typically utilized in GRMHD simulations 
\citep[e.g.,][and references therein]{Porth19}.
%%%
However, there is no guarantee that
the outer torus with the constant angular momentum 
considered is actually realized.
We also point out that the inner-edge of the torus
is typically placed quite 
close to the BH ($\sim$ a few $\times 10~r_{g}$)
simply due to a limitation of finite computational cost.
Furthermore, a new type of accretion via wind-fed process is 
proposed \citep[][]{ressler20a,ressler20b}. 
%%%
Thus, the value of  $\Omega_{F}$ may be affected by different physical states of the source (whichever is the torus or wind)
that supplies the magnetic field to the funnel region.

Third possibility is that
the jet base of M87 is anchored to the
innermost region of the accretion flow rather than threading the central Kerr BH. 
If this is the case, the suggested slow $\Omega_{F}$ is naturally explained.
%%%
However, it is fair to note that  highly accelerated jets reproduced in various 
GRMHD simulations are produced inside the funnel regions with high $\sigma$ values
which are anchored to  the central BH.
%%%
% In general, unbounded plasma from accretion flow is recognized as winds (e.g., Yuan and Narayan+)
% and it seems difficult to accelerate it up to relativistic speed since the 
% accretion flow is supposed to be magnetized as $\sigma<1$.
%%%
The boundary region in between the funnel region and the wind can be recognized 
 as a funnel-wall (FW) jet and it may be possible that the observed limb-brightening region  includes
a boundary layer zone between the pure funnel region and the disk-wind (funnel-wall),
where dissipation, turbulence generation and mass-loading may take place
\citep[][]{Hawley06}.
Interestingly, 
recent resistive GRMHD simulations show that 
plasmoids produced by magnetic reconnection have 
relativistic temperature \citep[][]{ripperda20,ripperda21}
that may trigger a relativistic flow. 
Therefore, the funnel-wall could produce 
relativistic blobs. 
Reconnection-driven particle acceleration in 
relativistic shear flows triggered by
Kelvin-Helmholtz instability 
could be another possibility
\citep[][]{sironi21}.
%%%
The VLBI data also show that fast moving blobs 
in the M87 jet are located at the limb brightening region 
\citep[e.g.,][]{Mertens16,Hada17,Park19}.
Thus, the FW jet with a slower $\Omega_{F}$ 
has a potential to mitigate the problem of 
the apparent discrepancy between the observed and the GRMHD-simulation-predicted velocity field profile, 
although further scrutiny should be definitely needed.

\subsection{Magnetic field strength on the event horizon}

\subsubsection{$B_{H}$ estimation by assuming BZ-process}

Here we discuss
magnetic field strength on the horizon scale
($B_{\rm H}$) 
by assuming BZ process is in action at the jet base of M87.
The essence of the BZ process is the energy extraction of the rotation energy of the BH
via magnetic fields anchored to the event horizon.
BZ process works under the condition 
$0< \Omega_{F}<\Omega_{H}$
\citep[see further recent discussions][]{king21,komissarov21}.
%%%%
%Since this condition naturally holds for a slow $ \Omega_{F}$,
%here we explore the field strength at the event horizon ($B_{\rm H}$) 
%by assuming that the BZ process is in action in M87. 
%%%
The  BZ power can be given by
\begin{align}
 L_{\rm BZ}\approx &&
7.5\times 10^{45}\chi_{-2} \frac{\Omega_{F}(\Omega_{H}-\Omega_{F})}{\Omega_{H}^{2}} \nonumber \\
&& 
\left(\frac{B_{H}}{10^{3}~{\rm G}}\right)^{2}
\left(\frac{r_{H}}{10^{15}~{\rm cm}}\right)^{2}
~{\rm erg~s^{-1}} ,
\end{align}
where  
$\chi$ is the geometrical factor given by
\begin{align}
\chi = \int^{\theta_{H}}_{0}   \sin^{3}\theta d\theta
=
\frac{2}{3} -\frac{3}{4}\cos\theta_{H} +\frac{1}{12}\cos3\theta_{H}.
\end{align}
\citep[e.g.,][]{Beskin00, takahashi21}
and $r_{H}$ is the outer horizon radius of the black hole
(see Appendix).
Here we normalized $\chi$ with a typical value 
suggested at the jet base 
$\chi_{-2}=\chi/10^{-2}$ \citep[e.g.,][]{Beskin00, Tchekhovskoy11}. 
We should bear in mind that 
there is uncertainty in the value of 
$\chi$ due to $\theta_{H}$.
%%%
Regarding the black hole spin, 
we assume the allowed range of the black hole spin as 
$0.5\le a/M_{\bullet} \le 1$
\citep[e.g.,][]{Zamaninasab14,Nakamura18}
since too small spin is not able to explain the required jet power.
%%%
By combining the estimated
$\Omega_{F}$ and assumed 
$\Omega_{H}$, 
here we will explore the range of
$0.02 \lesssim \Omega_{F}/\Omega_{H} \lesssim 0.1$.
\footnote{
The ratio of the angular velocity of the magnetic field line and the event horizon
can be rewritten as follows:
%%%%%%
%Since $\Omega_{F}=X_{F}/M_{\bullet}$ and $\Omega_{H}=
%a/2 M_{\bullet} r_{H}$ hold
%where $a$ is the spin parameter of the Kerr BH,  
%%%%%%%%
%The key quantity $\Omega_{F}/\Omega_{H}$ can be written by
%%%
$\frac{\Omega_{F}}{\Omega_{H}}
= 2\Omega_{F} M_{\bullet} \left(\frac{1+\sqrt{1-(a/M_{\bullet})^{2}}}{a/M_{\bullet}} \right)$.}
%%
%This clearly guarantees the mandatory condition for BZ process.
%%%%
%As was already emphasized, 
%the estimated  $\Omega_{F}/\Omega_{H}$ in 
%typical values in GRMHD simulations.
%%%%
In most of the literature, 
$ \Omega_{F}/\Omega_{H}\sim (0.2-0.5)$ are
considered 
\citep[e.g.,][]{McKinney06,Tchekhovskoy10,Penna13,Takahashi18} 
and our estimation is somewhat smaller than
those estimations.
Therefore,
higher $B_{H}$ will be needed to compensate 
to keep the total jet power.

Figure~\ref{fig:BZpower} presents the estimated range of
$B_{\rm H}$ in the allowed range of   $\Omega_{F}/\Omega_{H}$
together with 
the assumption of  $L_{BZ}\approx L_{\rm j}$
which holds unless significant dissipation happens during its propagation.
We conservatively allow a fairly wide 
range of the jet power as
$1\times 10^{42}~{\rm erg~s^{-1}}\lesssim  L_{\rm j} 
\lesssim 1 \times 10^{44}~{\rm erg~s^{-1}}$
\citep[e.g.,][]
{Reynolds96, Bicknell96, Owen00, Stawarz06,deGasperin12}.
%%%
Following the recent GRMHD simulations of highly magnetized jets
\citep[e.g.,][]{Porth19, Ripperda19},
we set the value of the magnetic field threading angle as 
$\theta_{H}= 1$ radian and it leads to $\chi = 0.18$
(see Appendix).
%%%
Then, the estimated $B_{H}$ lies in the range of 
$2\times 10^{2}~{\rm G}\lesssim  B_{\rm H} 
\lesssim  4\times 10^{3}~{\rm G}$,
which is comparable to the estimation by
\citet[][]{blandford19}.
This is larger than those estimated at the EHT photon ring region 
\citep[][]{EHT19_5, EHTC8, EHT_MWL21}.
%%
%We will further discuss it in the next sub-section.
%%%
The corresponding 
dimensionless magnetic flux on the event horizon scale ($\phi_{\rm BH}$)  
%%%
%\citep[][]{Tchekhovskoy11}
%%%
is estimated as
$\phi_{\rm BH}\equiv
(\Phi_{\rm BH}/\dot{M}r_{g}^{2}c)^{1/2} 
\approx 22
(B_{H}/10^{3}~{\rm G})
(\dot{M}/10^{-3}~{\rm M_{\odot}~yr^{-1}})^{-1/2}$
where 
$\Phi_{\rm BH}\approx B_{H} r_{g}^{2}$, and
$\dot{M}$ are the magnetic flux threading the black hole, 
and the mass accretion rate onto the black hole, respectively.
The mass accretion rate is adopted from \citep[][]{EHTC8}.
%%
%indicated by \citet[][]{Kuo14} is used for the normalization,
%the obtained $\phi_{\rm BH}$ in Eq.~(\ref{eq:phi_BH}) 
%becomes the lower limit.
%%%%%%
Accretion flows with $\phi_{\rm BH}\sim 1$
are classified as Standard and Normal Evolution
\citep[SANE:][]{Narayan12} state,
while the accretion flows with a larger $\phi_{\rm BH}$ such as
$\phi_{\rm BH}\gtrsim 15$
are conventionally referred as
Magnetically Arrested Disks
\citep[MAD:][]{Igumenshchev03, Narayan03, 
Tchekhovskoy11, EHT19_5} state.
%%%%%%%
The obtained $\phi_{\rm BH}$ obviously 
indicates that M87 is in a MAD regime.
%%%%%
%Our estimate of  $\phi_{\rm BH}$ is comparable
%with an estimation of
%for numerous blazar and radio galaxy samples 
%in \citet[][]{Zamaninasab14} and the 
%$\phi_{\rm BH}$ value
%adopted in the MAD model for M87 in \citet[][]{Chael19}.
%%%%%%
Although the estimation of the $B_{H}$ 
value depends on the adopted jet power, 
the estimated $\phi_{\rm BH}$ is 
consistent with the suggestion that M87 is MAD 
\citep[][]{EHTC8}.

%\subsection{memo of possibly remaining A/I}

%Possible reason of apparent contradiction with Takahashi et al. 2017should be discussed

%Estimate of the degree of the Doppler boosting at the jet base based on vphi

\subsubsection{Consistency check with EHT results}

%%%
%Radiation properties  a highly magnetized jet base
%is poorly constrained in GRMHD simulation studies
%due to floor density and thus 
%the radiation in this region is considered as 
%unreliable so that $\sigma$-cutoff is usually introduced 
%that cut off emissions from high-$\sigma$ jet 
%\citep[e.g.,][]{Moscibrodzka17,EHT19_5, Chael19}.
%Therefore, our methodology of constraining 
%$B_{H}$ assuming that BZ process is in action
%is one of the unique ways.
%%%%%%%
%Therefore, the careful comparison between 
%a lower limit 
%of B-field strength in your model and the upper limit in the EHTC model %(GRMHD model) 
%in the same (or similar) region, will be reasonable.  
%%%
Since the averaged magnetic field strength 
at the photon-ring region is estimated as
$B_{\rm ph-ring}\approx (5-30)~{\rm G}$
\citep[][]{EHT19_5,EHTC8,EHT_MWL21},
at a glance, one may concern that a large $B_{H}$ may 
emit excess synchrotron radiation that largely 
exceeds the observed photon-ring flux 
about 0.5~Jy at 230~GHz.
%%%
Therefore, it is worth to check 
whether the above estimated $B_{H}$ can be 
consistent with the EHT observation. 
%%%%%%%%
To this end, it is straightforward to
directly map   magnetic field strengths at the 
jet base using GRMHD+GRRT simulation data.
%%%%
%i.e., finding the discrepancy of the B-field strength between the region at the vicinity of the event horizon and the region significantly contributing to the flux detected by EHT. 
%%%
%Based on the suggestion of MAD state at the M87 jet base \citep[][]{EHTC8}, 
%%%%
In Figure~\ref{fig:EHTtest},
we show an example of the mapping of  magnetic field strength at the 
M87 jet base.
We find that a case of relatively dimmer jet 
can make it possible to realize a large $B_{H}$ 
without violating the EHT observational results
(see details for the model parameters in Appendix).
%%%
%This result indicates that investigating 
%physical processes of electron heating and acceleration is %essential for estimating $B_{H}$.
%%%
Further detailed comparisons between the model
and the observed images will be addressed as
in future work.

%%%
%It is well known 
%electron heating is more efficient for low $\beta$
%(Howes+, Rowan+17).
%%%%%%
%The two rows show simulations using different underlying models for electron heating: the top row (H10) uses the turbulent heating prescription of Howes (2010), and the bottom row (R17) uses the reconnection heating prescription of Rowan et al. (2017). The simulation data in both rows is averaged in time and azimuth.
%%%%%%

%The breakdown of charge-starved situation
%is supposed to be realized and it may
%essentially mitigate this problem.
%%%%%%% 
%If the electron temperature 
%is less than the canonical choice of 
%EHTC models, 
%then the overshoot may be relaxed.
%\footnote{
%Detailed studies using 
%two-temperature GRRMHD simulations 
%can explore this issue. 
%Indeed, it seems to show a 
%slightly lower temperature in the disk region 
%than that of the beta-prescription model as shown in EHT %paper 8.} 

\section{Summary}

Motivated by the measured velocity field profile of 
the M87 jet inthe KaVA large program
by \citet[][]{Park19}
that shows a slower acceleration compared 
to those suggested by GRMHD simulations,
we explore how to mitigate this apparent discrepancy
by using a semi-analytic SRMHD jet model proposed by \citet[][]{TT03}
consistently solving the trans-magnetic field structure.
%%%
We summarize our findings as follows:

\begin{itemize}

\item

By comparing TT03 model with the observed M87 jet velocity profile,
we find that the model can
reproduce the logarithmic feature of the velocity profile, 
and fit the observed data when choosing 
$c/(100r_{g}) \lesssim \Omega_{F} \lesssim c/(70r_{g})$
which is by a factor of 7-10
slower than the typical $\Omega_{F}$ in GRMHD simulations
\citep[e.g.,][]{McKinney06,Tchekhovskoy10}.
We discussed the possibility that
different boundary conditions lead to 
different values of $\Omega_{F}$.

\item
While a total specific energy (${\cal E}$) of each streamline changes the terminal bulk Lorentz factor,
a slower angular velocity of the magnetic fields  ($\Omega_{F}$) 
 makes a light-cylinder radius ($r_{\rm lc}$) larger  
and it consequently push out a starting point of the jet acceleration.
This provides us a new possibility to mitigate the apparent deviation 
between the KaVA observation of the M87 jet and 
GRMHD-simulation based prediction.

\item

%Since the central Kerr black hole with its angular velocity ($\Omega_{H}$) 
%in GRMHD simulations 
%needs to spin fast to produce the enough jet power,
%the condition $0<\Omega_{F}<\Omega_{H}$ always holds for 
%the obtained $\Omega_{F}$.
%%%%%%%
By assuming Blandford-Znajek (BZ) process is in action 
with the total jet power of $10^{42-44}~{\rm erg~s^{-1}}$,
we estimate the magnetic field strength 
on the event horizon scale in M87.
%(and using the field strength 
%estimated on the ISCO scale),
Then, 
it is estimated as
$2\times 10^{2}~{\rm G} \lesssim B_{H} 
\lesssim 5\times 10^{3}$~G 
for the total jet power  
of $10^{42-44}~{\rm erg~s^{-1}}$
in order to compensate for the effect of slower $\Omega_{F}$ than previously thought.
The corresponding $\phi_{\rm BH}$
suggests that M87 is in a MAD regime.
This is similar to the argument 
by \citet[][]{blandford19}
claiming the need of the spinning of the hole together with
the magnetic field of order of $\sim 10^{3}$ G 
to launch the M87 jet.

\item

It is important to note that 
this work only discusses the extreme cases where dissipation 
does not work merely for simplicity.
%%%
Although the simplification is basically 
justified to some extent \citep[e.g.,][]{Chatterjee19},
it may be possible to have co-existence of
a feeble dissipation at the jet base
\citep[e.g.,][]{ripperda20, ripperda21, sironi21}.
%%%
Inclusion of the magnetic reconnection process
would probably facilitate to explain  
the observed characteristic limb brightening structure
observed at 86~GHz  \citep[][]{Hada16,Kim18}.
%%%
One of the ways to test this scenario
would be to probe $u_{p}$ at the base of deeper jets ($\lesssim 10^{2}~r_{g}$)
with VLBI observations of high spatial resolution and compare the model prediction by \citet[][]{takahashi21} and those future VLBI observations.

\end{itemize}

%%%%%%%%%%%%%%%%%%%%%%%%%%%%%%%%%%%%%%%%%%%%%%%%%%

\bigskip
\leftline{\bf \large Acknowledgment}
\medskip

\noindent

We thank the referee for the comments that helped improve the overall clarity of the manuscript.
We thank 
the EHT Collaboration internal reviewer Y. Mizuno, 
who carefully checked the manuscript and
provided constructive comments.
We are grateful to M. Nakamura and K. Toma 
for fruitful discussions and useful comments.
We also thank K. Ohsuga and H. R. Takahashi for discussions on GRMHD simulations.
This work was partially supported by
the MEXT/JSPS KAKENHI
(JP18H03721,
JP17K05439, 
JP18K1359,
JP21H01137, and
JP22H00157).
 J.P. acknowledges financial support through the EACOA Fellowship awarded by the East Asia Core Observatories Association, which consists of the Academia Sinica Institute of Astronomy and Astrophysics, the National Astronomical Observatory of Japan, Center for Astronomical Mega-Science, Chinese Academy of Sciences, and the Korea Astronomy and Space Science Institute.
 This research was also supported by MEXT as “Program for Promoting Researches on the Supercomputer Fugaku” (Toward a unified view of the universe: from large scale structures to planets, JPMXP1020200109) and JICFuS.

%%%%%%%%%%%%%%%%%%%%%%%%%%%%%%%%%%%%%%%%%%%%%%%%%%

\appendix
In this Appendix, a brief note on
the estimation of BZ power is presented
where the background metric is written by
Boyer-Lindquist coordinates.
The electromagnetic energy flux from the horizon is generally
given by  $T^{\mu\nu}k_{\nu}=T^{r}_{0}$ where  
$T^{\mu\nu}$ is the stress-energy tensor, and 
$k_{\nu}$ is the time-like Killing vector.
For axisymmetric case,
the radial component of the electromagnetic energy flux 
is explicitly given by 
\citep[][]{BZ77,Znajek77}
\begin{eqnarray}
T^{r}_{0}&=&
- \epsilon_{0} 
\frac{\Omega_{F} B_{\phi}}{\Sigma_{H}\sin\theta} \Psi_{\theta} \nonumber  \\ 
&=&
2\epsilon_{0}M_{\bullet}r_{H}
\Omega_{F}(\Omega_{H}-\Omega_{F}) 
\left(\frac{\Psi_{\theta} }{\Sigma_{H}}\right)^{2}, 
\end{eqnarray}
where
$a=J/M_{\bullet} =2M_{\bullet}\Omega_{H} r_{H}$,
$J=M_{\bullet} \Omega_{H}r_{H}^{2}$, 
$r_{H}=M_{\bullet}+\sqrt{M_{\bullet}^{2}-a^{2}}$, and
$\Omega_{H}$
are, 
the spin parameter,
the angular momentum,
the radius of the outer horizon, and
the angular velocity of the Kerr BH, respectively.
We also note that $\Sigma_{H}=r_{H}^{2}+a^{2}\cos^{2}\theta$
and 
the toroidal magnetic fields on the event horizon ($B_{\phi}$) is given by
$B_{\phi}=\frac{[\Omega_{F} (r_{H}^{2}+a^{2})-a] \sin\theta }{\Sigma_{H} } A_{\phi,\theta}$.
%%%
In this Appendix,
we locally use the conventionally used
magnetic flux function of
$\Psi = 
\Psi_{\rm BH}\left(\frac{r}{M_{\bullet}}\right)^{p}(1-\cos\theta)$
where $p$ is the power-law index describing the field geometry and $\Psi_{\rm BH}$ is constant.
The case of $p=0$ corresponds to the conical magnetic field (split monopole),
while  $p=1$ describes the parabolic one.
%%%%

The net BZ power in the jet is obtained by integrating the 
EM energy  flux evaluated at the event horizon, where 
magnetic fields are within the half opening angle of the foot-point 
of the black hole magnetosphere $\theta_{H}$.
Then, the BZ power is given by
\begin{eqnarray}
L_{\rm BZ}
&=&
\int^{2\pi}_{0}d\phi  \int^{\theta_{H}}_{0}  T^{r}_{0}\Sigma_{H} \sin\theta d\theta  \nonumber \\
&=&
2\pi \epsilon_{0} \left(\frac{r_{H}}{M_{\bullet}}\right)^{2p} \Psi_{\rm BH}^{2}
\int^{\theta_{H}}_{0}   \frac{2M_{\bullet}r_{H}}{\Sigma_{H}} \Omega_{F} (\Omega_{H}-\Omega_{F})
 \sin^{3}\theta d\theta        . 
\end{eqnarray}
Thus we obtain $L_{\rm BZ} \propto \Omega_{F}(\Omega_{H}-\Omega_{F})  
\Psi_{\rm BH}^{2}$
and hence
a slow $\Omega_{F}$ leads to a smaller  
$L_{\rm BZ}$
\citep[e.g.,][]{Tchekhovskoy08, Beskin00}.
Then the BZ power can be further written as
\begin{eqnarray}
L _{\rm BZ}
\approx 
2\pi \epsilon_{0}\Psi_{\rm BH}^{2} 
\left(\frac{r_{H}}{M_{\bullet}}\right)^{2p}
\left(\frac{a}{2M_{\bullet}r_{H}}\right)^{2}
\frac{\Omega_{F}(\Omega_{H}-\Omega_{F})}{\Omega_{H}^{2}}\chi .
\end{eqnarray}
When the magnetic field geometry is
split-monopole for instance, 
the BZ power is estimated as 
\begin{eqnarray}
 L_{\rm BZ}
&\approx& 
7.5\times 10^{45}
\chi_{-2}
\frac{\Omega_{F}(\Omega_{H}-\Omega_{F})}{\Omega_{H}^{2}}
%\left(\frac{M_{\bullet}}{10^{9}~M_{\odot}}\right)^{2} 
\left(\frac{B_{H}}{10^{3}~{\rm G}}\right)^{2}
\left(\frac{r_{H}}{10^{15}~{\rm cm}}\right)^{2}
~{\rm erg~s^{-1}} ,
\end{eqnarray}
where $\Psi_{\rm BH}
\approx  B_{H}r_{H}^{2}$ is the magnetic flux on the horizon
and here we omit $a$ dependence in $r_{H}$ merely for simplicity.
The case of parabolic magnetic field geometry
needs to multiply the additional factor 
of $r_{H}/M_{\bullet}=1+\sqrt{1-(a/M_{\bullet})^{2}}$, which maximally becomes the factor of 2 at most.

\appendix
In this Appendix, we explain
of one example of GRMHD simulation
presented in the discussion in details.
%%%%%%%%%%%%%%%%%%%%
% kawashima-sentences
%%%%%%%%%%%%%%%%%%%%
We used a GRMHD simulation snapshot in semi-MAD state with normalized spin parameter $a_* = 0.9375$, which is same with that shown in \citep{kawashima21b} performed by using a GR(-Radiation)MHD simulation code \texttt{UWABAMI} \citep{Takahashi16}.
%%%%%%
To set the electron temperature $T_{\rm e}$,
we assumed an $R$-$\beta$ prescription
 \citep[e.g.,][]{moscibdodzka16, EHT19_5}
 given by
\begin{eqnarray}
\frac{T_{p}}{T_{e}}= 
R_{\rm high}\frac{\beta^{2}}{1+\beta^{2}}+
R_{\rm low}\frac{1}{1+\beta^{2}}
\end{eqnarray}
where 
$T_{\rm p}$ and $\beta$ are
the proton temperature and plasma beta 
(i.e., the ratio of gas pressure to magnetic pressure),
respectively.
The term
$R_{\rm high}$ parameterizes $T_{e}$
in a high-$\beta$ accretion flow region, 
while $R_{\rm low}$  parameterizes $T_{e}$
in a low-$\beta$ jet region.
%%%%%%%%%%%%%%%%%%%%%%
By performing GRRT calculations with \texttt{RAIKOU} code \citep[][]{kawashima19,kawashima21b},
we search for combinations of $R_{low}$ and  $B_{H}$  
that satisfy the total flux about 0.5~Jy at 230~GHz\citep[][]{EHT19_1}
and $B_{H}$ estimated in the present work.
Following EHTC work,
we conservatively choose $\sigma_{\rm cut}=1$ to fully
exclude the emission coming from the density floor region.
We set 
$R_{\rm high} = 160$ and $R_{\rm low} = 2$ 
where $R_{\rm low}$ is set to be slightly larger 
than that used in the most of works $R_{\rm low} = 1$. 
The $R_{\rm low}$ larger than unity will be applicable, since the radiative cooling via synchrotron and inverse-Compton scattering processes can reduce the electron temperature and one temperature assumption in the low-$\beta$ region can breakdown \citep[see, e.g., Figure 17 in][]{EHTC8}.
%%%%%%%%%%%%%%%%%%%%%%%%%%%
In this work, we tried  $R_{\rm low}$ than unity to examine the possibility of a large $B_{\rm H}$.
This is because a larger mass accretion rate is required to reproduce the same radiative flux when $R_{\rm low}$ is larger (i.e., lower temperature in the highly magnetized region.). The magnetic field strength is proportional to the root square of the mass accretion rate in GRMHD simulations, i.e., the stronger magnetic field appears when we set a larger $R_{\rm low}$ with keeping reproducing the observed radiative flux.
%%%%%%%%%%%%%%%%%%%%%%%%%%%%%%%%%%%%%%%%%%%%%%%%%%%%%
%For simplicity, 
%non-thermal emissions are not included here.
%%%%%%%%%%%%%%%%%%%%%%%%%%
A large $R_{\rm low}$
reduces the 230~GHz flux density from the jet
and  thus can avoid the overshooting of 
the observed EHT flux density.
%%%%%
%\citep[][]{EHTC8} indeed demonstrates
%the comparison of $R_{\rm low}= 1$ and 
%$R_{\rm low}= 10$ and confirm the reduce the 
%flux density for $R_{\rm low}= 10$.
%%%%%%
%in order to realize 0.5~Jy 230~GHz flux.
%%%
%Since the synchrotron emission from the jet is more dominant than
%that from the accretion flow in this case,
%photon ring images only weakly depends on $R_{\rm high}$. 
%Therefore, we fix $R_{\rm high}=40$

%%%%%%%%

%\newpage
%\setcounter{page}{1}
%\bibliographystyle{apj}
%\bibliography{ken}

\footnotesize
%%%%%%%%%%%%%%%%%%%%%%%%%%%%%% References:
%\bibliography{ref.bib,aeireferences.bib,GRRT.bib,GRRT2.bib}{}
%\bibliography{GRRT.bib,GRRT2.bib}{}
%\bibliography{GRRT.bib}{}
%\bibliography{kino21.bib}{}

\bibliography{ms.bbl}{}

%%%%%%%%%%%%%%%%%%%%%%%%%%
%%%%%%%%%%%%%%%%%%%%%%%%%%

%%%%%%%%%%%%%%%%%%%%%%%%%%
%%%%%%%%%%%%%%%%%%%%%%%%%

%%%%%%%%%%%%%%%%%%%%%%%%%%
%%%%%%%%%%%%%%%%%%%%%%%%%
%%%%%%%%%%%%%%%%%%%%%%%%%%
%%%%%%%%%%%%%%%%%%%%%%%%
%%%%%%%%%%%%%%%%%%%%%%%%%%
%%%%%%%%%%%%%%%%%%%%%%%%%%
%%%%%%%%%%%%%%%%%%%%%%%%%%
%%%%%%%%%%%%%%%%%%%%%%%%%
%%%%%%%%%%%%%%%%%%%%%%%%%%
%%%%%%%%%%%%%%%%%%%%%%%%%
%%%%%%%%%%%%%%%%%%%%%%%%%%
\begin{figure} 
\includegraphics
[width=18cm]
{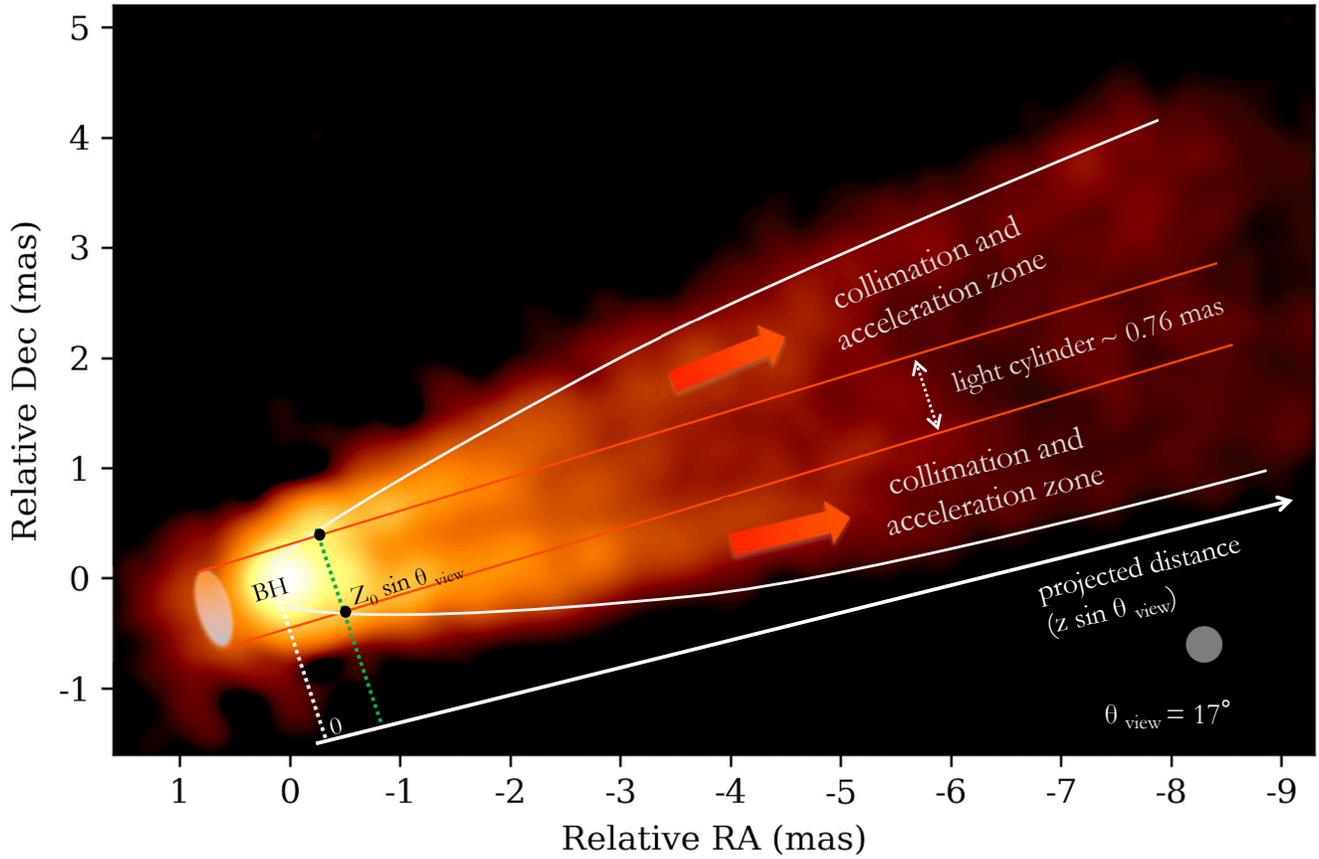}
\caption
{Schematic illustration 
of the model overlaid on stacked M87 image 
at 43~GHz (Cui et al. in preparation).
TT03 model is applicable to the region outside the light cylinder.
The model applicable range starts from
$z\ge Z_{0}$, at which the outer boundary wall
and the light cylinder have intersection points,
and it is located at $\sim (2-4)\times 100~r_{g}$
from the central BH.
As explained in the main body,
the semi-parabolic boundary wall is determined by
the jet-width measurement by VLBI
\citep[][]{Asada12, Hada13}.
}
\label{fig:model-obs}
\end{figure}
%%%%%%%%%%%%%%%%%%%%%%%%%%
%%%%%%%%%%%%%%%%%%%%%%%%%%
\begin{figure} 
\includegraphics
[width=20cm]
{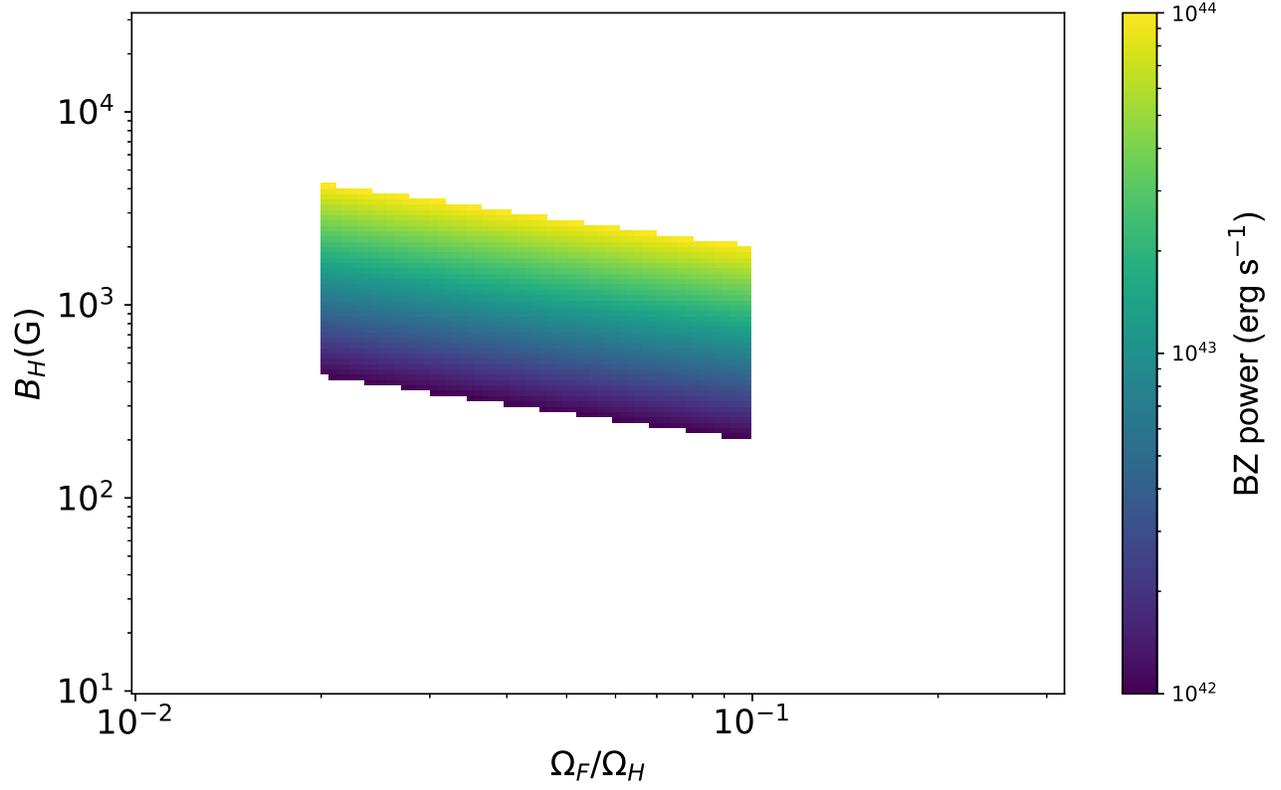}
\caption
{Estimate of  the magnetic field strength 
threading the event horizon. 
The horizontal axis shows the ratio $\Omega_{F}/\Omega_{H}$
where $\Omega_{F}$ is constrained as Eq.~(\ref{eq:omega_f}). 
The estimate here is simply derived by equating 
$L_{\rm BZ}$ (Appendix) to the total jet power in M87
suggested as 
$1\times 10^{42}~{\rm erg~s^{-1}}\lesssim  L_{\rm j} 
\lesssim 1 \times 10^{44}~{\rm erg~s^{-1}}$. The color bar
shows the corresponding $L_{\rm BZ}$.}
\label{fig:BZpower}
\end{figure}
%%%%%%%%%%%%%%%%%%%%%%%%%%
%%%%%%%%%%%%%%%%%%%%%%%%%%
\begin{figure} 
\includegraphics
[width=15cm]
%{Bmag_2side01200.pdf}
{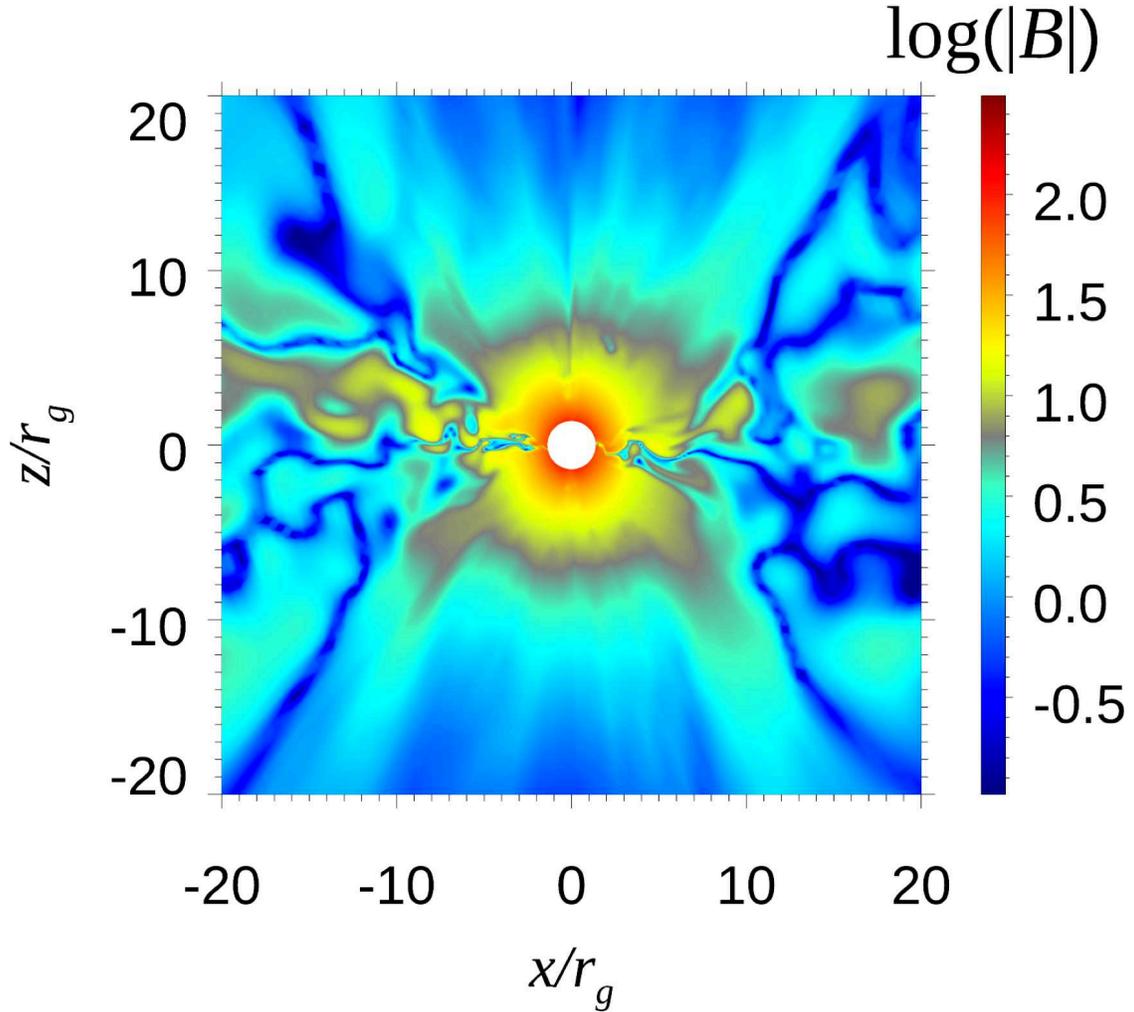}
\centering
\caption
{
Map of magnetic field strength of a GRMHD simulation in logarithmic scale in a poloidal plane.
Since GRMHD simulations are scale free, we normalized the mass accretion rate by carrying out GRRT calculations with \texttt{RAIKOU} code \citep[][]{kawashima19,kawashima21b} using the parameters for M87* ($M_{\bullet} = 6.5\times 10^{9} M_{\odot}$ and the distance $D = 16.9$ Mpc in such a way that the resulting image reproduces the ring like image with the radiative flux $\sim 0.5$ Jy at 230 GHz observed by EHT \citep{EHT19_1}.
%%%%%%%%%%
Here we demonstrate the case of
%$B_{\rm H} = 195$ G in the GRMHD simulation model
$B_{\rm H} = 91$ G in the GRMHD simulation model
This is roughly consistent with the $B_{\rm H} \gtrsim 200$ G estimated from our analytical jet model.
Although the GRMHD simulation model shown here is in the semi-MAD state, we also confirmed that the MAD state achieved $B_{\rm H} \gtrsim 200$ G with the same procedure (Kawashima et al. in prep.)
Although a detailed comparison with the EHT images is beyond the scope of this paper, 
it is clear that the reproduction of the photon-ring with 
the synchrotron flux density about 0.5 Jy at 230~GHz is feasible.}
\label{fig:EHTtest}
\end{figure}

\end{document}